\documentclass[a4paper,10pt]{article}
\usepackage{amsmath,amsfonts}
\usepackage{graphicx}
\newcommand{\blankline}{\vskip .3cm}
\newcommand{\beq}{\begin{equation}}
\newcommand{\eeq}{\end{equation}}
\newcommand{\matrice}{\begin{pmatrix}}
\newcommand{\ematrice}{\end{pmatrix}}

\let\a=\alpha \let\b=\beta  \let\g=\gamma  \let\d=\delta
     \let\th=\theta  \let\l=\lambda
\let\m=\mu    \let\n=\nu         \let\p=\pi    
\let\s=\sigma      
    
\let\G=\Gamma \let\D=\Delta   \let\L=\Lambda 
         
  \let\ee=\epsilon

\let\==\equiv
\let\io=\infty
\def\ie{{i.e. }}

\def\TT{{\cal T}}\def\NN{{\cal N}}

\def\to{\rightarrow}
\def\la{\left\langle}
\def\ra{\right\rangle}

\def\Tr{{\rm Tr}\,}
\def\la{\langle}
\def\ra{\rangle}

\newcommand{\bea}{\begin{eqnarray}}
\newcommand{\eea}{\end{eqnarray}}

\begin{document}

{\normalsize \hfill ITP-UU-07/19}\\
\vspace{-1.5cm}
{\normalsize \hfill SPIN-07/12}\\
${}$\\
\vspace{3cm}

\centerline{\LARGE (2+1)-Dimensional Quantum Gravity as the Continuum Limit of} \blankline \centerline{\LARGE
Causal Dynamical Triangulations} \blankline
\vspace{1cm}
\centerline{D. Benedetti${}^a$, R. Loll${}^a$, F.Zamponi${}^b$}
\blankline
\blankline \blankline \centerline{${}^a$\it Spinoza Institute and Institute for Theoretical Physics,}
\centerline{\it Utrecht University,}
\centerline{\it Leuvenlaan 4, NL-3584 CE Utrecht, The Netherlands}
\blankline \centerline{${}^b$\it Service de Physique Th\'eorique,}
\centerline{\it  Orme des Merisiers, CEA Saclay,}
\centerline{\it F-91191 Gif-sur-Yvette Cedex, France}
\blankline \blankline %\centerline{24/04/2007} \blankline
\blankline\blankline \blankline

\vspace{1cm}

\centerline{ABSTRACT}
\vspace{.2cm}

\noindent{We perform a non-perturbative sum over geometries
in a (2+1)-dimensional quantum gravity model given in terms of Causal
Dynamical Triangulations. Inspired by the concept of triangulations of
product type introduced previously, we impose an additional notion of
order on the discrete, causal geometries. This simplifies the combinatorial
problem of counting geometries just enough to enable us to calculate
the transfer matrix between boundary states labelled by the area of the
spatial universe, as well as the corresponding quantum Hamiltonian of
the continuum theory. This is the first time in dimension
larger than two that a Hamiltonian has been derived from such a model
by mainly analytical means, and opens the way for a better understanding
of scaling and renormalization issues.
}

\eject

\section{Introduction}

Trying to come up with {\it any} method by which the non-perturbative
regime of four-dimensional quantum gravity could be probed
quantitatively is a challenging task. Given the strongly interacting
nature of the problem, one would hope to be able to attack it by a
combination of numerical and analytic tools,
informing and complementing each other. Starting in two spacetime
dimensions \cite{2d-ldt,3d-ldt}, a potentially powerful approach to
quantum gravity has been developed over the last few years
under the name of {\it Causal Dynamical Triangulations}, or CDT for
short. In the grand tradition of {\it Quantum Regge Calculus} and the program of
(non-causal, Euclidean) {\it Dynamical Triangulations}, its aim is
the evaluation of a non-perturbative path integral over spacetime
geometries, represented as a state sum over piecewise flat
triangulations (see \cite{reggereview, 4drev, scratch} for review
material).

Although CDT models for pure gravity
can be solved exactly in a variety of ways in (1+1) dimensions
\cite{2d-ldt,2dlessons,difra1,2deulor,difra-calogero,topo1,topo2},
it is difficult to extend any of the analytical treatments to the case of
higher dimensions, where most of the important
results so far have come from numerical simulations
\cite{ajl-prl,semi,spectral,universe}.\footnote{CDT's precursor of
``Dynamical Triangulations" did see some interesting attempts
at using analytic tools to understand the model's phase structure in four
dimensions, at both strong and weak coupling \cite{dt1,dt2}.}
An exception to this is the attempt to solve the (2+1)-dimensional CDT
model\footnote{More precisely, a version of CDT defined on an extended
ensemble of spacetimes allowing for certain ``wormhole" degeneracies
of the local geometry.}
by mapping ``thick slices" of its spacetime geometries to configurations of
a two-dimensional matrix model with ABAB-interaction \cite{ABAB1,ABAB2,ABAB3}.
An exact solution of the matrix model relevant to
computing the CDT transfer matrix has been
given \cite{zinn-just-1,zinn-just-2}, but its rather complicated and implicit form has so far
been an obstacle to performing its continuum limit analytically, and thus
extracting a quantum Hamiltonian.

This situation is not particularly surprising, as most known solvable statistical
models are only one- or two-dimensional.
Even a simple spin model like the Ising model has been solved only in two dimensions,
and for zero external field.
One can therefore already anticipate that an extension of analytical methods and
results to the case of higher-dimensional CDT models will be a challenging task.
One possible way to approach this problem is to try to modify the model
by making simplifying assumptions to improve its solubility, but without changing its
physical content in the continuum.
The main aim of the present work is to study such a modified model of
three-dimensional CDT, which is a particular case of a class of models
introduced in \cite{bianca-bh}.\footnote{One of the main motivations there was
the desire to formulate CDT models with (approximate) spherical symmetry,
like those describing a black hole.}
Compared with the full CDT model, the ensemble of geometries we will
use possesses an additional notion of ``order",
which will enable us to bring a number of technical tools to bear on its
solution. As we will see, this leads to a non-trivial dynamics for the
three-dimensional quantum universe, justifying the ansatz retrospectively.

Despite the technical simplification it introduces,
the particular three-dimensional CDT model we will be considering
is still too complicated to be solved in full generality.
In this situation, the special properties of three-dimensional continuum
gravity come to our help. Since there are no local, gravitonic field
excitations in three dimensions, the dynamical content of this model
is encoded in a number of {\it global} metric variables describing
the universe as a whole.
One can perform a classical, canonical constraint analysis \`a la Dirac
of (2+1)-dimensional general relativity to see explicitly how
the number of degrees of freedom is reduced from infinite to finite
(see, for example, \cite{carlip-3d}).

The way in which we will make use of this property in the non-perturbative
path-integral setting of the CDT approach, which works entirely in terms of
geometries (that is, on the quotient space of metrics modulo diffeomorphisms),
is by integrating over almost all degrees of freedom in the transfer matrix of
our model. The
resulting ``effective" quantum Hamiltonian we are able to derive in the continuum
limit will therefore only depend on those degrees of
freedom we {\it anticipate} to be left over as the physically relevant ones. Most importantly,
this is the volume (two-dimensional area) of the spatial universe.
An extension to
include an additional, global Teichm\"uller parameter may also be technically
feasible, but will not be presented here. In the course of the calculation,
we find more evidence for a non-perturbative
cancellation of the divergent conformal mode of the Euclideanized gravitational
path integral\footnote{Also a continuum treatment in proper-time gauge,
which presumably comes closest to the coordinate-free CDT formulation,
strongly suggests that the kinetic term
for the conformal factor of the spatial geometries (the would-be propagating
field degree of freedom)
is cancelled in the path integral by a Faddeev-Popov determinant in the
measure \cite{dasgupta-loll}.},
already evident in other investigations of
causal dynamically triangulated models in dimensions three \cite{numeric-k}
and four \cite{universe}.

This paper is organized as follows. In Sec.~\ref{sec-model} we describe
the geometric configurations whose partition function we are going
to study and use as a transfer matrix. The configurations are from an ensemble of
triangulations of product type,
interpolating between successive constant-time slices, with free boundary conditions and
boundary cosmological constants, conjugate to the areas of the slices.
In Sec.~\ref{sec-inversion} we introduce the first tool that will help us in solving the
three-dimensional CDT model,
the inversion formula for {\it heaps of pieces}.
In Sec.~\ref{sec-matrices} we reformulate the partition function in terms of the transfer
matrices for a set of one-dimensional hard bi-coloured dimer models.
In Sec.~\ref{sec-zeros} we study the location of the singularities of the partition function
and define a critical surface in the parameter space, which is the boundary of the
region where the model is convergent and well defined.
In Sec.~\ref{sec-replica} we introduce the second tool we will need for the solution,
the replica trick for products of random matrices.
This enables us to identify a critical point on the critical surface
where the mean volume goes to infinity, thus making it a potential candidate for
taking a non-trivial continuum limit.
We can give an exact solution to the model at this critical point, and
approximate it to the desired order in the displacement parameter when moving away from it.
In Sec.~\ref{sec-continuum} we start exploring the continuum properties of the
model by deriving expectation values for the volume and curvature of spacetime.
In Sec.~\ref{sec-gluing} we construct the area-to-area transfer matrix by summing
over unphysical degrees of freedom, and
explain the procedure for extracting information about the spatial volume of the universe.
In Sec.~\ref{sec-hamiltonian} we show that for a vanishing bare inverse gravitational constant,
one obtains a well-defined quantum Hamiltonian acting on the Hilbert space of the
area eigenstates. By contrast, a canonical scaling ansatz for the gravitational constant
does not seem to lead to a meaningful result.
Finally, our conclusions are presented in
Sec.~\ref{sec-conclusions}. --
Four appendices collect various technical results and discussions needed in the main text.

\section{Introducing the model} \label{sec-model}

\subsection{The product type triangulations}

Part of our strategy for trying to solve the non-perturbative three-dimensional
model of quantum gravity defined by causal dynamical triangulations (CDT) is to
identify a suitable (sub-)class of all CDT configurations whose superposition
can be tackled {\it analytically}. The aim of such a reduction is to simplify the
analytic treatment, without
eliminating relevant degrees of freedom to such a degree that the universality
class is changed, compared to that of the full CDT model.\footnote{This assumes
that the full CDT model, which has been studied in Monte Carlo simulations \cite{numeric-k},
leads to an essentially unique three-dimensional quantum gravity theory.
Strictly speaking, rather little is known in dimension three about the universality
classes of statistical models of random geometry like the one we are using.
There are certainly more, but we are only interested in those which possess a
continuum interpretation in terms
of quantum gravity.} To achieve this, we propose to
work with a set of triangulated, causal geometries which have an additional
``order" imposed on them. This order is mild in the sense of
not implying ``isometries" of the triangulations: both the local spacetime curvature
and the local curvature of spatial slices can still vary arbitrarily. Our model
therefore has less order than the hexagon model considered in \cite{bianca-hex}, which has
flat spatial slices.

The way in which we will introduce more structure or order on triangulated
spacetimes -- inspired by similar earlier ideas \cite{difra2,bianca-bh} -- is that of applying
the building principle inherent in {\it causal} dynamical triangulations twice over.
This ``causality principle" recognizes that in a metric spacetime of Lorentzian
signature not all directions
are equivalent, but there is a distinguished (class of) time direction(s), in line with
the existence of light cones and causal relations, none of which are present
in spaces of purely Euclidean signature. In CDT this causal structure is
implemented via a discrete global time slicing \cite{3d-ldt}.
This simply means that each simplicial
building block of a triangulated spacetime must be contained in exactly one
spacetime ``sandwich" (the region between discrete proper times $t$ and $t+1$).
This implies a causal ordering on the simplices, without constraining
the local curvature degrees of freedom.

In addition to this physically motivated choice of a distinguished time direction,
which is one of the key ingredients of the approach of causal dynamical
triangulations,
we will introduce here a second distinguished direction, but this time purely
for computational convenience, and under the assumption that it will not
affect the universal properties of the gravitational model\footnote{An assumption
that obviously will have to be borne out by the final result.}. This will roughly
speaking correspond to an additional slicing, but this time in one of the spatial
directions. As explained in \cite{bianca-bh}, the resulting structure can be thought of
as a staggered fibration in a piecewise flat setting.

\begin{figure}[ht]
\centering
\vspace*{13pt}
\includegraphics[width=9cm]{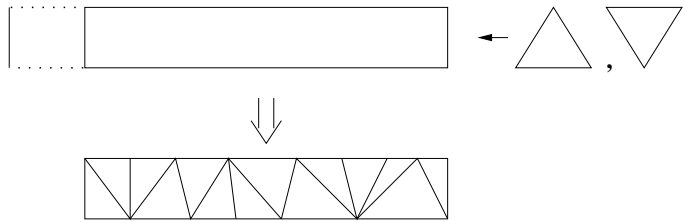}
\vspace*{13pt}
\caption{\footnotesize A triangulated strip constructed as a tower over a one-dimensional link.}
\label{segment-tower}
\end{figure}

To illustrate the main idea behind this type of triangulation let us start
from the easier (1+1)-dimensional CDT model. A spacetime contributing
to the sum over geometries is usually described as a sequence of
triangulated strips glued together, where each strip represents a
piece of spacetime between proper times $t$ and $t+1$, and the
lengths of the spatial boundaries of adjacent strips must match pairwise.
For our purposes, it is useful to think of a strip as being constructed
``sideways" (see Fig.\ref{segment-tower}): starting from a segment (a time-like link),
build a frame on it and then fill the frame with some sequence of up- and down-triangles.
We will refer to this construction as
``building a two-dimensional {\it tower} over a link". The entire two-dimensional
triangulated spacetime can then be regarded as a fibration over a one-dimensional
chain of links.

\begin{figure}[ht]
\centering
\vspace*{13pt}
\includegraphics[width=9cm]{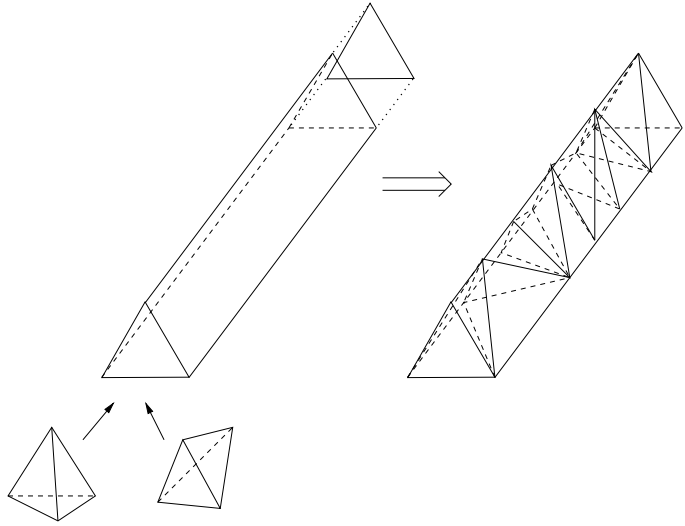}
\vspace*{13pt}
\caption{\footnotesize A triangulated prism constructed as a tower over a
two-dimensional triangle.}
\label{triangle-tower}
\end{figure}

More generally, {\it product triangulations} \cite{bianca-bh} are obtained by building towers
over (arrays of) higher-dimensional simplices, such as triangles. How a three-dimensional
tower is built over a triangle is illustrated in Fig.\ref{triangle-tower}.\footnote{It should be noted that
the figures do not represent faithfully the {\it metric} properties of the building blocks, that is, their edge
lengths. By construction, up to a relative factor between space- and time-like links, the edge
lengths in a causal dynamical triangulation are all identical.}
If the base space is not just a single triangle, but a two-dimensional triangulation
consisting of $n$ triangles, we can construct a three-dimensional product
triangulation by erecting a tower over each of them, in such a way that the boundary
triangulations of the resulting prisms again match pairwise.
A general $(n+m)$-dimensional product triangulation is a simplicial manifold
constructed by consistently building $(k+m)$-dimensional towers (for all $k\leq n$) over the
$k$-dimensional sub-simplices of an $n$-dimensional {\it base} simplicial manifold.

In the present work we will use this construction in the context of (2+1)-dimensional
causal dynamical triangulations. By definition, these spacetimes can be regarded
as product triangulations whose base space -- like in (1+1) dimensions -- is a
one-dimensional triangulation in the time direction, and the fibres over each
link of the base space are the sandwiches between integer values of the discrete
time $t$ introduced earlier. Now, instead of allowing sandwich geometries which
are arbitrary triangulations of thickness $\Delta t=1$, we will impose an additional
product structure on them. Namely, they should themselves have the form of a
sequence of up- and down prisms, as depicted in Fig.\ \ref{tower-sandwich} below.
That is, in addition to the slicing of the entire spacetime
corresponding to the global discrete proper time, each
sandwich possesses a discrete slicing in a given spatial direction. The number of
slices (that is, the number of prisms in a sandwich) is allowed to vary from
sandwich to sandwich.

An equivalent way of characterizing such spacetimes is to regard them as fibrations
over a two-dimensional base space (with one space- and one time-dimension)
which is itself an arbitrary (1+1)-dimensional CDT, and where over each triangle of
the base space we have erected a tower ``filled" with a sequence of tetrahedra, in such
a way that the triangulations on the faces of neighbouring towers match and can be
glued together consistently. In this way, we may think of our triangulations as being
(1+1+1)-dimensional.

\subsection{The partition function}\label{sec-part}

We will concentrate on the dynamics of a ``sandwich geometry",
given by the transition amplitude from a spatial geometry at time $t$ to
one at time $t+1$, in other words, the transfer matrix $\hat T$ of the
causal dynamical triangulation (CDT) model, which in turn contains
information about its quantum Hamiltonian $\hat H$
by virtue of the relation $\hat T=e^{-a \hat H}$ (see \cite{3d-ldt} for details).

Since pure gravity in (2+1) dimensions does not possess any local,
propagating degrees of freedom, we expect that most details of the
spatial geometries will be dynamically irrelevant, leaving only
the spatial two-volume $A(t)$ and the Teichm\"{u}ller parameters (together
with their canonically conjugate momenta) as
physical degrees of freedom. Therefore, rather than calculating the
matrix elements
\beq\label{firsttransfer}
G(g_1,g_2,\Delta t=1)=\langle g_2|\hat T |g_1\rangle
\eeq
of the transfer matrix
from an arbitrary spatial triangulated geometry $|g_1\rangle$ at time $t$ to an
arbitrary $|g_2\rangle$ at time $t+1$, we will only keep track of the two
boundary areas $A_1$ and $A_2$ and evaluate the reduced matrix elements
\beq\label{area-area}
  G(A_1,A_2,\Delta t=1)=
  \langle A_2|\hat T |A_1\rangle,
\eeq
where $|A_i\rangle$ is a normalized linear combination of
states $|g_i\rangle$ with given area $A_i$ (see
\cite{ABAB1} for a detailed discussion). Note that one is still summing over the
{\it same} sandwich geometries as in (\ref{firsttransfer}), so the reduced matrix
elements still capture the effective dynamics of {\it all} geometric excitations of
the sandwich geometry.\footnote{The presence of physical degrees of freedom
beyond the spatial area depends on the topology of the spatial slices. There are
none for the case of spherical slices considered in \cite{numeric-k,ABAB1}, and there is one
Teichm\"uller parameter
in the present model, where we choose boundary conditions corresponding to
cylindrical spatial slices \cite{polchinski}. Work is under way to determine whether the
calculation of matrix elements presented here is still feasible when the
dependence on this parameter is kept explicitly.}
Once these matrix elements are known, a next step is to try to extract
the continuum Hamiltonian operator $\hat{H}$ of the system from an expansion
in the short-distance cutoff $a$ as $a\rightarrow 0$ according to
\beq \label{HfromT}
  \langle A_2|\hat T |A_1\rangle =
  \langle A_2|e^{-a \hat{H}} |A_1\rangle =
  \langle A_2|\left(\hat{1}-a\hat{H}+O(a^2)\right)|A_1\rangle.
\eeq

We will denote by $N_{ij}$ the number of simplices having $i$ vertices
on the initial boundary spatial slice at time $t$ and $j$ on the final one at
$t+1$.
In this way $A_1$ (resp. $A_2$) is given by $N_{31}$ (resp. $N_{13}$).
The prescription for evaluating the discrete one-step propagator (\ref{area-area}) is
given by
\beq\label{area-area-cdt}
  G(N_{31},N_{13},\Delta t=1)=\sum_{\TT_{|N_{31},N_{13}}}\frac{1}{C_{\TT}}\ e^{-S_{EH}},
\eeq
where the sum is over all sandwich triangulations $\TT$
with fixed boundary areas $N_{31}$ and $N_{13}$,
$S_{EH}$ the Wick-rotated discrete Einstein-Hilbert
action, and $C_{\TT}$ the order of the automorphism group of $\TT$.
Taking into account boundary terms in the action to ensure the
correct propagator behaviour, we find
\beq\label{}
  S_{EH}=\alpha (N_{13}+N_{31}) + \beta N_{22} +\gamma N
\eeq
as an explicit expression for the gravitational action, where $N$ is the number of triangle
towers in the sandwich, and where we have introduced the parameters
\beq\label{alphabeta}
\begin{split}
  &\alpha =-c_1 k + b_1 \lambda \\
  &\beta = c_2 k + b_2 \lambda \\
  &\gamma = c_3 k
\end{split}
\eeq
depending on the dimensionless bare cosmological and inverse Newton constants
$\lambda$ and $k$, and on positive numerical constants $c_i$ and $b_i$ characterizing
the geometric construction (see appendix \ref{App-action}).

As one might have anticipated from previous investigations of related
three-dimensional quantum gravity models,
the evaluation of (\ref{area-area-cdt}) remains a challenging task, also for
our specific choice of ensemble of triangulations.
Our main aim will be to calculate its discrete Laplace transform
\beq\label{laplace}
\begin{split}
  Z(x,y,\Delta t=1)&=\sum_{N_{31}}\sum_{N_{13}}x^{N_{31}}y^{N_{13}}G(N_{31},N_{13},\Delta t=1)\\
  &=\sum_{N_{31}}\sum_{N_{13}}(x e^{-\alpha})^{N_{31}}(y e^{-\alpha})^{N_{13}}\sum_{\TT_{|N_{13},N_{31}}}
  e^{-\beta N_{22}-\gamma N} \\
&=\sum_{\TT} u^{\frac{N_{31}}{2}}v^{\frac{N_{13}}{2}} w^{N_{22}} e^{-\g N}  ,
\end{split}
\eeq
where the sum runs over all triangulations $\TT$ compatible with the structure discussed above,
and we have introduced the weights $\sqrt{u}=x e^{-\alpha}$, $\sqrt{v}=y e^{-\alpha}$
and $w= e^{-\beta}$.
In writing (\ref{laplace}), we have set $C_{\TT}$ equal to one, because for the
chosen boundary conditions its contribution in the continuum limit will be negligible.
The function $Z(x,y,\D t=1)$
can be thought of either as the generating function of $G(N_{31},N_{13},\Delta t=1)$,
or as the partition function of the sandwich geometry with free boundary conditions
(thus summing over all values of the boundary volume) and with additional
weights $x=e^{-\lambda_{in}}$ and $y=e^{-\lambda_{out}}$, corresponding to
additional boundary cosmological terms in the action.

To get back the transfer matrix from $Z(x,y,\Delta t=1)$, one needs to keep $x$
and $y$ distinct and variable. However, as shown by other examples
(see the end of section 3.1 for the two-dimensional case, and \cite{numeric-k}
for a numerical illustration in three dimensions), one may be able to extract
non-trivial information about the phase structure of the model by considering
the special values $x=y=1$, which simplifies the evaluation of~(\ref{laplace}).

\begin{figure}[ht]
\centering
\vspace*{13pt}
\includegraphics[width=9cm]{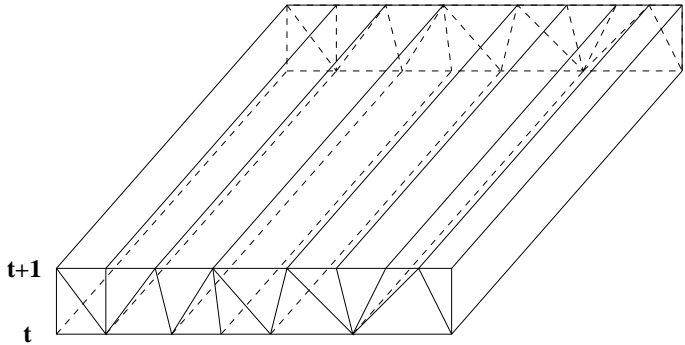}
\vspace*{13pt}
\caption{\footnotesize A ``sandwich geometry" of product type in the
(1+1+1)-dimensional CDT model, built over a
given triangulated base strip. We split the calculation of its partition function into
two parts, calculating first the partition function for such a geometry
and then summing over all possible sequences of towers. By ``towers of type
(2,1)" we will mean the towers built on a triangle with two vertices at time $t$
and one at time $t+1$, and similarly for ``towers of type (1,2)".
Each of the towers is filled up with a sequence of tetrahedra,
as illustrated in Fig. \ref{triangle-tower}. There are three different possibilities
for how a tetrahedron can be oriented inside one of the prisms (see \cite{bianca-bh}
for more geometric details). The apparent ``straightness" of the towers is again an
artefact of our representation, which emphasizes the product structure.}
\label{tower-sandwich}
\end{figure}

At this stage we will exploit the special product structure of our chosen
spacetime geometries to split the sum over all sandwich triangulations into two simpler
sums. The idea is to first perform the sum over all fillings (i.e. triangulations)
of towers for a {\it fixed} triangulated base strip, and then
to perform the sum over all possible triangulations of the base strip (see
Fig.\ref{tower-sandwich}), schematically
\beq \label{sum-split}
  \sum_{\TT_{sandwich}}=\sum_{\TT_{base \ strip}}\hspace {.3cm}\sum_{\TT_{towers}}.
\eeq

We are thus naturally led to studying the partition function $Z_N$ for
sandwich geometries of product type with $N$ towers, and investigating its
properties in the large-$N$ limit. In this sense we are treating the second
distinguished direction, which defines the slicing within the sandwich
geometry, in the same way as we treat the time direction in (1+1)-dimensional CDT.
It is related to (\ref{laplace}) via
\beq \label{zndefine}
 Z(x,y,\Delta t=1)=\sum_{N}  e^{-\gamma N} Z_{N}(u,v,w),
 \eeq
with $Z_{N}$ expressible as
\beq \label{Z_N}
   Z_{N}(u,v,w) = \sum_{S_N}  Z_{S_N}(u,v,w)=
    \sum_{S_N} \sum_{\TT_{S_N}}
  u^{\frac{N_{31}}{2}} v^{\frac{N_{13}}{2}} w^{N_{22}},
\eeq
where the sequences $S_N$ consist of $N$ prism towers,
$N-R$ of them of type (2,1) and $R$ of type (1,2), for any $R<N$.

\section{Inversion formula} \label{sec-inversion}

The special product form of our sandwich geometries will enable us to make
use of an {\it inversion formula}, relating its partition function to that of a
model of one dimension less. As we will review in the next subsection, this
technique essentially provides a complete solution to the CDT model in
dimension (1+1). This is not true in the much more involved case of
CDT in dimension (1+1+1) we are addressing here, but -- as we will
demonstrate -- it will nevertheless allow
us to make substantial progress in the evaluation of the partition function.

\subsection{The two-dimensional case}

The partition function of the full (1+1)-dimensional CDT model \cite{2d-ldt},
more specifically, the partition function of its equivalent dual formulation in terms of
three-valent graphs \cite{difra1}, can be expressed as the
inverse of the partition function $Z^{hd}$ of a hard-dimer model in {\it one} dimension
\cite{difra2}, namely,
\beq \label{fund2}
  Z^{2d}_t(u=g^2)= \frac{1}{Z_t^{hd}(-u)},
\eeq
where
\beq\label{hd}
  Z_t^{hd}(u)= \sum_{{\rm hard}\ {\rm dimer}\ {\rm config.}\ D} u^{|D|},
\eeq
$g= e^{-\lambda}$ is the volume weight assigned to each triangle
and $t$ is the number of time steps.
We recall the proof of this inversion formula in appendix \ref{App-inversion}.

This is not an isolated result, but can be seen as a particular instance of a
more general mathematical result which can already be found in earlier work
\cite{viennot}, where also the notion of a {\it heap of pieces}
was introduced. Roughly speaking, a heap of pieces is a partially ordered set
whose elements occupy
columns out of a finite set of columns, and where direct-order relations only exist
between elements in the same column or in neighbouring ones.
It is not difficult to realize that triangulations of product type are closely
related to the heap-of-pieces construction, and in particular that the structure
of the (1+1)-dimensional
CDT model (in its dual picture) is that of a heap of dimers.

One can also re-express the partition function (\ref{hd}) in a transfer matrix formulation.
This is done by introducing a two-dimensional vector space associated to each site,
with the vector $(1,0)$ corresponding to the empty state, and $(0,1)$ to
an occupied dimer state at the site. If we attach a weight 1 to the
transition empty-empty (between neighbouring dimers), a weight
$\sqrt{u}$ to the transition dimer-empty or empty-dimer, and a weight 0 to the
transition dimer-dimer, we obtain a transfer matrix $M$ associated with such
a transition. Since the dimer partition function
for the case of periodic boundary conditions is given by $\Tr (M^t)$, we can rewrite
(\ref{fund2}) as
\beq\label{theta2}
  Z^{2d}_t(u=g^2)=\frac{1}{\Tr \begin{pmatrix} 1 & {\rm i} \sqrt{u} \cr
{\rm i} \sqrt{u} & 0\cr\end{pmatrix}^t}.
\eeq
Since $\Tr(M^t)=\lambda_+^t+\lambda_-^t$, where $\lambda_{\pm}=(1\pm\sqrt{1-4u})/2$
are the two eigenvalues of the transfer matrix, we have
\beq \label{Z-2d}
  Z^{2d}_t(u=g^2)\sim \frac{1}{\lambda_+(u)^t}
\eeq
for large $t$, which is non-analytical in $u=1/4$, the critical point.

Alternatively, this critical point can already be found by analyzing
the one-step propagator
\beq\label{zstrip}
  Z^{2d}_{t=1}(i,j \vert g)=\sum_{strips\ \TT_{|i,j}} \frac{1}{C_{\TT}} e^{-S_{EH}}
\eeq
of the (1+1)-dimensional model, given by a sum over triangulated strips
$\TT_{i,j}$ with an initial boundary of length $i$ and a final boundary of
length $j$. Since the Einstein-Hilbert action consists only of the
volume term (multiplied by the cosmological constant $\lambda$),
the propagator (\ref{zstrip}) reduces to
\beq\label{}
  Z^{2d}_{t=1}(i,j \vert g=e^{-\lambda})=e^{-\lambda (i+j)}\sum_{strips\ \TT_{|i,j}} 1
\eeq
with the corresponding generating function
\beq\label{gen2}
  Z^{2d}_{t=1}(x,y \vert g=e^{-\lambda})=\sum_{i,j\geq 0} x^i y^j e^{-\lambda (i+j)}
  \sum_{strips\ \TT_{|i,j}} 1.
\eeq
With the help of the inversion formula (and
considering $x$ and $y$ as the square roots of the weights of the
boundary dimers), the latter can now be rewritten as
\beq\label{gen2bis}
  Z^{2d}_{t=1}(x,y \vert u=e^{-2\lambda})=\frac{1}{\matrice 1 & {\rm i} y\ematrice
  \matrice 1 & {\rm i} \sqrt{u} \\
  {\rm i} \sqrt{u} & 0\ematrice \matrice 1 \cr {\rm i} x \ematrice}
  =\frac{1}{1-\sqrt{u}(x+y)}.
\eeq
We observe that by setting $x=y=1$ ($i.e.$ considering the partition
function for $\Delta t=1$ with free boundary conditions), we get a
singularity in the point $\sqrt{u}=\frac{1}{2}$. This is exactly the critical point,
and the continuum limit has to be taken by fine-tuning $\lambda$ to
its critical value $\lambda_c=\ln{2}$. It is straightforward to see that
in this limit $\langle V\rangle=-\frac{1}{Z}\frac{\partial Z}{\partial\lambda}$
diverges, as required. Of course, the full one-step propagator
(\ref{gen2bis}) with $x\neq y$ contains much more information; indeed, as
demonstrated in \cite{2d-ldt}, one can recover
from it the quantum Hamiltonian of the model in the continuum limit.

\subsection{The three-dimensional case}

In order to apply the inversion formula to the case of three-dimensional
causal triangulations, we will use the fact that the propagator of this
model can be characterized in terms of geometric data which are ``almost"
two-dimensional, an observation already underlying earlier investigations
of three-dimensional CDT in
terms of the ABAB-matrix model \cite{ABAB1,ABAB2,ABAB3}. The argument runs as follows.
Consider a three-dimensional sandwich geometry, made up of a single
layer of tetrahedra
glued together pairwise along their time-like triangular faces.
For a given tetrahedron, there are three possibilities for its orientation
relative to the initial slice at time $t$ and the final slice at time $t+1$,
characterized again by the numbers (i,j) of vertices it shares with either
slice.

\begin{figure}[ht]
\centering
\vspace*{13pt}
\includegraphics[width=11cm]{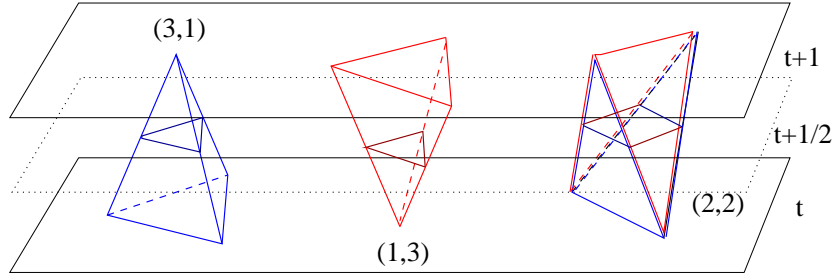}
\vspace*{13pt}
\caption{\footnotesize The three types of tetrahedral building blocks and their
intersections at time $t+1/2$.}
\label{blocks}
\end{figure}

Now take a two-dimensional spatial slice through the sandwich geometry
at time $t+1/2$. This is a well-defined prescription since the discrete time
$t$ can be continued uniquely inside the simplices by virtue of their flatness
\cite{bianca-bh}.
The surface at $t+1/2$ is piecewise flat, consisting of flat triangles
(coming from tetrahedra of type (3,1) and (1,3)) and flat squares
(from the tetrahedra of type (2,2)), see Fig.\ \ref{blocks}.\footnote{By a slight
abuse of language, we will keep referring to this geometry as a
``triangulation".} However, it is
clear that given this triangulation alone, we cannot
reconstruct the original three-dimensional sandwich
geometry. For this we need to know whether a given triangle at
$t+1/2$ came from a (3,1)- or a (1,3)-tetrahedron, or, equivalently,
whether a given link of the triangulation came from a time-like
(2,1)- or (1,2)-triangle. One way to keep track of this information is to
colour-code the two types of links, as shown in Fig.\ \ref{blocks}.
By saying that the one-step propagator cannot be characterized by
two-dimensional geometric data alone,
we meant the need for this additional colouring information.

\begin{figure}[ht]
\centering
\vspace*{13pt}
\includegraphics[width=9cm]{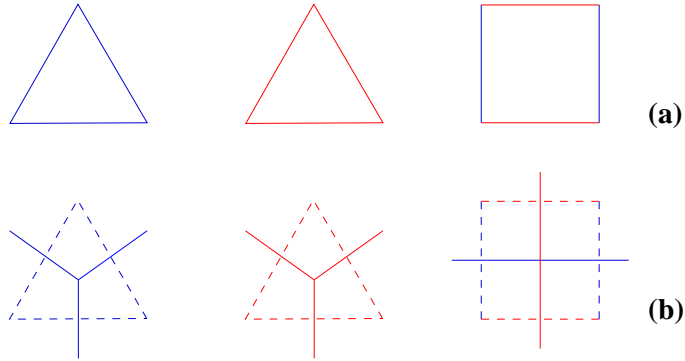}
\vspace*{13pt}
\caption{\footnotesize The elementary building blocks of the coloured
``triangulation" at time $t+1/2$ characterizing a sandwich geometry
(a), together with their duals (b).}
\label{vertices}
\end{figure}

In order to map the partition function for the bi-coloured two-dimensional
geometries to that of a hard-dimer model, we first go to the bi-coloured
graphs dual to the triangulations. The links of the dual graphs are obtained
by connecting the centres of neighbouring triangles and squares, and
their colour is inherited from the colour of the crossing link of the original
triangulation (see Fig.\ \ref{vertices}).

\begin{figure}[ht]
\centering
\vspace*{13pt}
\includegraphics[width=12cm]{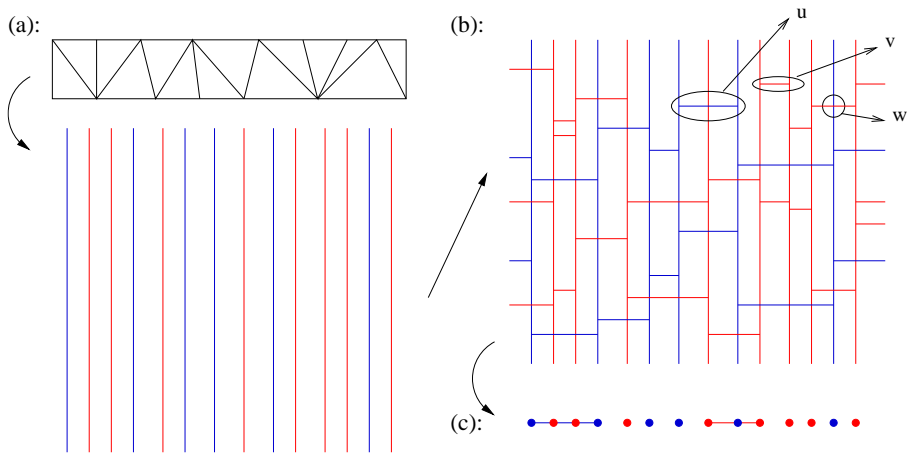}
\vspace*{13pt}
\caption{\footnotesize From sandwich geometries with product structure to hard dimers:
(a) vertical lines representing towers over a triangulated base strip; (b) horizontal links
encode how the tetrahedra are connected, the corresponding weights can be read off
as indicated; (c) projection to a sequence of hard-dimer configurations, as
explained in appendix A-II.}
\label{3d-dual}
\end{figure}

As a consequence of the product structure for the sandwich geometries,
also the dual graphs obtained in this way have a special form.
Namely, part of the dual graph can be drawn as
a fixed sequence of straight lines (which we will call
``vertical lines", and which are drawn vertically in Fig.\ \ref{3d-dual})
coming from the prism towers, blue for a tower over
a type-(2,1) triangle and red for a tower over a type-(1,2) triangle.
The sequence of blue and red towers
depends on the triangulation of the base strip.
In addition, the dual graph contains ``horizontal links", which
start on a vertical line of the same colour and end on the next
(to the right, say) vertical line of the same
colour. The position of the horizontal links encodes the three-dimensional
triangulation of the original sandwich geometries.

Looking at the subgraphs of one colour in the dual graph example of
Fig.\ \ref{3d-dual}, one notes that they are particular examples of
dual graphs for (1+1)-dimensional CDT spacetimes. It is indeed the
case that the blue graph is precisely the graph dual to the triangulation
of the initial two-dimensional surface of the sandwich geometry, and
the red graph the dual to its final surface. We may therefore think of
the bi-coloured graph as a particular superposition of these two
uni-coloured graphs. Bi-coloured graphs of this type are again of the
form of {\it heaps of pieces} in the sense of \cite{viennot},
which means that we can map them to dimer configurations as
in the (1+1)-dimensional situation above (c.f. appendix \ref{App-inversion}),
but with the difference that now the one-dimensional
dimer model will also be bi-coloured, with a fixed sequence of
blue and red sites and with the blue (red) dimers linking a blue
(red) site with the next one of the same colour, as depicted in Fig.\ \ref{3d-dual}.

The upshot of our considerations so far is that for a sandwich
geometry associated with a fixed triangulation $S_N$ of the base strip
an analogue of formula (\ref{fund2}) holds, namely,
\beq\label{fund3}
  Z_{S_{N}}(u,v,w)= \frac{ 1}{Z_{S_{N}}^{hcd}(-u,-v,w)},
\eeq
where
\beq\label{hcd}
  Z_{S_{N}}^{hcd}(u,v,w)= \sum_{ {\rm coloured}\ {\rm hard-dimer}\ {\rm config.}\ D_{|S_N}}
  u^{|D|_b}v^{|D|_r}w^{|\cap D|}
\eeq
is the partition function of the coloured dimer problem for a given
sequence $S_N$ of blue and red sites, $|D|_b$ ($|D|_r$) is the
number of blue (red) dimers in the configuration $D$,
and $|\cap D|$ is the number of crossings between dimers and sites of different colour.
We should emphasize that our application of the inversion formula
relates only to the {\it second} sum in (\ref{sum-split}), since we sum over
the triangulation of the towers for a {\it fixed} triangulation of the base strip.
This allows us to rewrite (\ref{zndefine}) as
\beq
\label{Zinv}
  Z(x,y,\Delta t=1) =\sum_N e^{-\g N} Z_N(u,v,w) = \sum_N e^{-\g N} \sum_{S_{N}}\hspace {.3cm}
  \frac{ 1}{Z_{S_{N}}^{hcd}(-u,-v,w)}, \hspace {.2cm}
\eeq
whose further evaluation will be the subject of the remainder of the paper.

\section{Random matrix formulation} \label{sec-matrices}

The coloured-dimer partition function can
be written in terms of transfer matrices, by introducing
three vectors (1,0,0), (0,1,0) and (0,0,1) corresponding
to an empty link, a blue link and a red link between neighbouring sites.
A transfer matrix can be associated with transitions
between these states, taking into account that hard dimers
are not allowed to touch each other, and that a dimer of a given
colour connects only sites of the same colour which are first
neighbours (among the sites with same colour), and may cross
sites of the opposite colour in between. The transfer matrices
can therefore be associated with the sites or vertices of the
dimer model, and their explicit form depends on the colour label
of the site.
The transition empty-empty gets a weight $1$, the transitions
dimer-empty or empty-dimer get a weight $\sqrt{u}$ or $\sqrt{v}$,
depending on the colour, and the crossing of a site by a
dimer of any colour gets a weight $w$; all other possibilities have
weight $0$.
In this way we can associate a matrix
\beq\label{A}
  \tilde A=\matrice 1 & \sqrt{u} & 0 \cr
                    \sqrt{u} & 0 & 0 \cr
                    0 & 0 & w \ematrice
\eeq
with every blue site, and a matrix
\beq\label{B}
  \tilde B=\matrice 1 & 0 & \sqrt{v} \cr
                    0 & w & 0 \cr
                    \sqrt{v} & 0 & 0 \ematrice
\eeq
with every red site.
Consequently, the partition function $Z_{S_{N}}^{hcd}(u,v,w)$
can be expressed as a product of $N$ matrices $\tilde A$
and $\tilde B$, ordered according to the sequence
specified by $S_{N}$ (note that the matrices
$\tilde A$ and $\tilde B$ do not commute).
We can now write the partition function (\ref{fund3}) associated with a fixed
sequence $S_{N}$
as\footnote{Our choice of taking the trace in the denominator implies periodic
boundary conditions in one of the spatial directions, by gluing together the first
and last prism tower of the sandwich. Since we are leaving the other spatial
direction open, the spatial slices of our model have the topology of a cylinder.
Instead of compactifying one of the directions, we could also have left it
open, in which case we would have to specify
the boundary lengths as boundary conditions, or introduce conjugate boundary variables
analogous to the $x$- and $y$-variables in (1+1) dimensions, eqs.\ (\ref{gen2}) and (\ref{gen2bis}),
and contract the product of matrices with the vectors
$(1\;ix_1\;ix_2)$ and $(1\;iy_1\;iy_2)$.
This would complicate the treatment of the partition function considerably.}
\beq\label{fund3-AB}
  Z_{S_{N}}(u,v,w)= \frac{ 1}{\Tr(ABAABABBB...)_{S_{N}}}
\eeq
where the matrices $A$ and $B$ are the same as $\tilde A$ and $\tilde B$, but
with the substitution $\sqrt{u}\rightarrow {\rm i}\sqrt{u}$,  $\sqrt{v}\rightarrow {\rm i}\sqrt{v}$.
In order to express the sum over sequences $S_N$ in eq.\ (\ref{Z_N}), it
is convenient to combine the two matrices and define a general transfer matrix
\beq\label{M_q}
  M_{q_j}(u,v,w)=q_j A + (1-q_j) B =\matrice 1 & q_j {\rm i}\sqrt{u} & (1-q_j){\rm i}\sqrt{v} \\
                       q_j{\rm i}\sqrt{u} & (1-q_j)w & 0 \\
                       (1-q_j){\rm i}\sqrt{v} & 0 & q_j w \ematrice,
\eeq
with $q_j$ taking the values 1 ($M=A$) or 0 ($M=B$). The sum over sequences $S_N$
is then replaced by a sum over sequences in
$\{0, 1\}^N$. In this formulation, the calculation of the partition
function assumes the form of a problem of products of random matrices
\cite{crisanti}, and we can rewrite the full partition function (\ref{zndefine}) as
\beq
\label{ZZZ}
  Z(x,y,\Delta t=1) =\sum_{N} e^{-\g N} \sum_{\{q_j\}_{N}} \frac{ 1}{\Tr\prod_{j=1}^N M_{q_j}(u,v,w)}.
\eeq
This in turn can be thought of as the average of (\ref{fund3-AB}) over
all possible configurations $\{q_j\}_N$, with
$q_j=1$ and $q_j=0$ each having probability $p=1/2$.
Introducing the notation
\beq
|P_{N,\{q\} }(u,v,w)|=\Tr\prod_{j=1}^N M_{q_j}(u,v,w),
\eeq
we can finally write
\beq\label{Z}
 Z(x,y,\Delta t=1) =\sum_N e^{-\g N}Z_N(u,v,w) = \sum_N (2 e^{-\g})^N \left\langle \frac{1}{|P_{N,\{q\} }(u,v,w)|}
  \right\rangle ,
\eeq
where we have used the normalized ensemble average
$\langle \bullet \rangle \equiv 2^{-N} \sum_{\{q_j\}} \bullet$.

\section{The zeros of the denominators} \label{sec-zeros}

An important feature to be noted about the inversion formula is that it maps an infinite
series with positive coefficients to the inverse of a finite sum with alternating sign
coefficients. The latter will have a real root corresponding to the radius of convergence
of the infinite series. This is evident in the (1+1)-dimensional case, formula (\ref{gen2bis}).
The same is clearly true for the (1+1+1)-dimensional case (\ref{fund3}) too, which will
have a two-dimensional {\it locus} of zeros in the three-dimensional parameter space
spanned by $u$, $v$ and $w$.
However, the location of this locus will depend on the sequence $\{ q_j\}$, and
we must keep in mind that we still have to sum over all the sequences $\{ q_j\}$.
Consequently, we will be interested in determining the envelope of all the loci.
In the absence of an analytical solution, which appears difficult to come by,
we will determine its location in the limit of infinite $N$ by studying
particular classes of sequences and by numerical computations, in conjunction
with a bit of (well-motivated) conjecture.

\subsection{u = v}

Let us temporarily assume that $x=y$ (and eventually $=1$) in the Laplace transform of
the one-step propagator, (\ref{laplace}), and therefore $u=v$.
This clearly makes things easier, but should still permit us to extract information
about the phase structure of the model, as is the case in both two and
three dimensions (see \cite{ABAB1} and \cite{ABAB2} for illustrations of the latter).
This is not an implausible proposition in the sense that we want to find the
critical line in the $k$-$\lambda$ plane, which concerns the ``bulk" behaviour
of the spacetime and should be insensitive to details of the boundary
data, like those encoded in $x$ and $y$.

In order to determine the zeros of the polynomials
$|P_N(u,v=u,w)|=\Tr\prod_{j=1}^N M_{q_j}(u,v=u,w)$,
we write this product as $A^{n_1}B^{n_2}...A^{n_{M-1}}B^{n_M}$ for some sequence
of $M\leq N$ positive integers $\{n_1,...,n_M\}$ which sum up to $N$, $\sum_{i=1}^Mn_i=N$.
Note that from $u=v$ follows $B=J A J$, where the matrix
\beq
  J=\matrice 1 & 0 & 0\\ 0 & 0 & 1\\ 0 & 1 & 0\ematrice
\eeq
is a projector ($i.e.$ $J^2=Id$), so that we can write
\beq \label{AJ^M}
  |P_{N,\{q\} }|=\Tr(A^{n_1}J A^{n_2}J...A^{n_{M-1}}J A^{n_M}J).
\eeq
Note furthermore that
\beq
  A^n=\matrice (\alpha^n)_{11} & (\alpha^n)_{12} & 0 \cr
                    (\alpha^n)_{12} & (\alpha^n)_{22} & 0 \cr
                    0 & 0 & w^n \ematrice ,
\eeq
where $\alpha^n$ is the n-th power of the matrix
\beq
\alpha :=
\matrice 1 & {\rm i} \sqrt{u} \\
{\rm i} \sqrt{u} & 0\ematrice
\eeq
found in (\ref{theta2}) above, and which has the following properties:
\beq \label{a^n_11}
  (\alpha^n)_{11}=\frac{\lambda_+^{n+1}-\lambda_-^{n+1}}{\lambda_+-\lambda_-}
  =u^{\frac{n}{2}} U_n(\frac{1}{2 \sqrt{u}})
\eeq
\beq
  (\alpha^n)_{12}=i\sqrt{u}(\alpha^{n-1})_{11}
\eeq
\beq
  (\alpha^n)_{22}=(\alpha^n)_{11}-(\alpha^{n-1})_{11}
\eeq
where $\lambda_{\pm}=(1\pm\sqrt{1-4u})/2$ are the eigenvalues
of the matrix $\alpha$ ($w$ is the third eigenvalue of the matrix $A$),
and $U_n(x)$ is the n-th Chebyshev polynomial of the second kind.

\subsubsection{u=0}
For the subcase $u=0$ everything is trivial since $|P_N(0,0,w)|=1$ for every sequence.
(We exclude degenerate sequences with $N$ matrices of one type and none of the other, which
would yield $|P_N(0,0,w)|=1+w^N$. At any rate, they would give a negligible contribution
in the average.) We thus have $Z_N(0,w)=1$.

\subsubsection{w=0}
For the subcase $w=0$ the trace (\ref{AJ^M}) reduces to
\beq\label{w=0}
  |P_{N,\{q\} }|=z_0=\prod_{i=1}^{M} (\alpha^{n_i})_{11}\equiv
 \prod_{i=1}^{M}
  u^{\frac{n_i}{2}} U_{n_i}(\frac{1}{2 \sqrt{u}})
\eeq
by virtue of (\ref{a^n_11}), which implies
that (\ref{AJ^M}) becomes the product of $M$ uncoupled
partition functions of the simple (uncoloured) hard-dimer problem.

Note first of all that the term $u^{\frac{n}{2}}$ in (\ref{a^n_11}) does not vanish at $u=0$,
because its product with $U_n(\frac{1}{2 \sqrt{u}})$ gives a polynomial of order $[\frac{n}{2}]$
in $u$ whose zeroth order term is $1$.
From mathematical handbooks we know that $U_n(x)$ has zeros only in the interval $[-1,1]$,
more precisely, at the values $x_i=\cos{(\frac{i}{n+1}\pi)}$, for $i=1,...,n$. It follows
that no zeros can occur for $u<1/4$. By contrast, for fixed $u>1/4$ there is always
a sequence, for some $N$, such that $|P_{N,\{q\} }|$ has zeros between $1/4$ and $u$.
For positive $\epsilon=u-\frac{1}{4}$, the occurrence of the first root
is when $\frac{1}{2\sqrt{\frac{1}{4}+\epsilon}}\leq x_1=\cos{(\frac{1}{n+1}\pi)}$.
For small $\epsilon$ this requires a large $n$. And this in turn means
that (some but not necessarily all of) the sequences contributing to the critical
behaviour at $w=0$ and
$u=\frac{1}{4}$ possess a large group of consecutive matrices of
the same kind; particular examples of this are configurations like $A B^{N-1}$.
A predominance of such configurations would signal a decoupling of
neighbouring slices in the three-dimensional geometry, and a geometric
degeneracy of the model there.

\subsubsection{u,w$\neq$0}

\begin{figure}[h]
\begin{center}
$\begin{array}{c@{\hspace{1in}}c}
\multicolumn{1}{l}{} &
    \multicolumn{1}{l}{} \\ [-0.53cm]
\includegraphics[width=4cm]{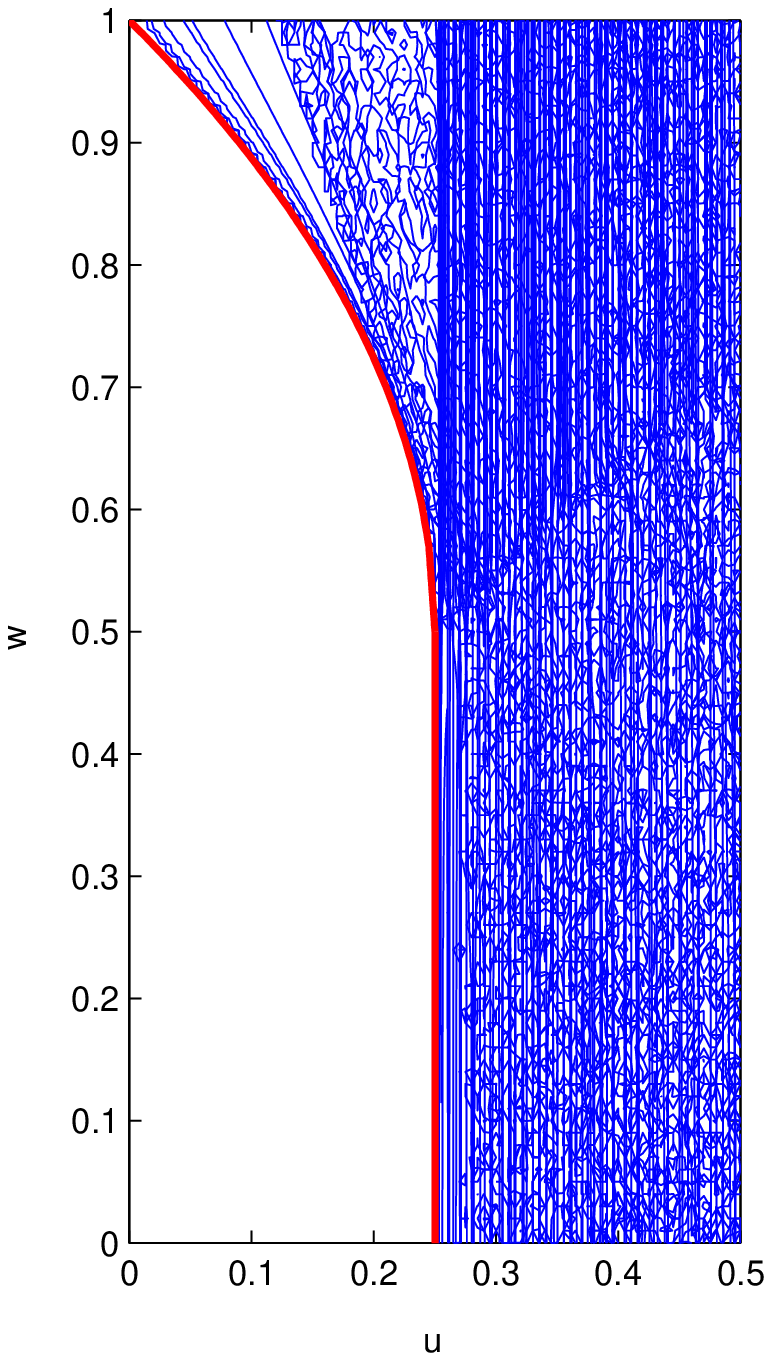} &
\includegraphics[width=7cm]{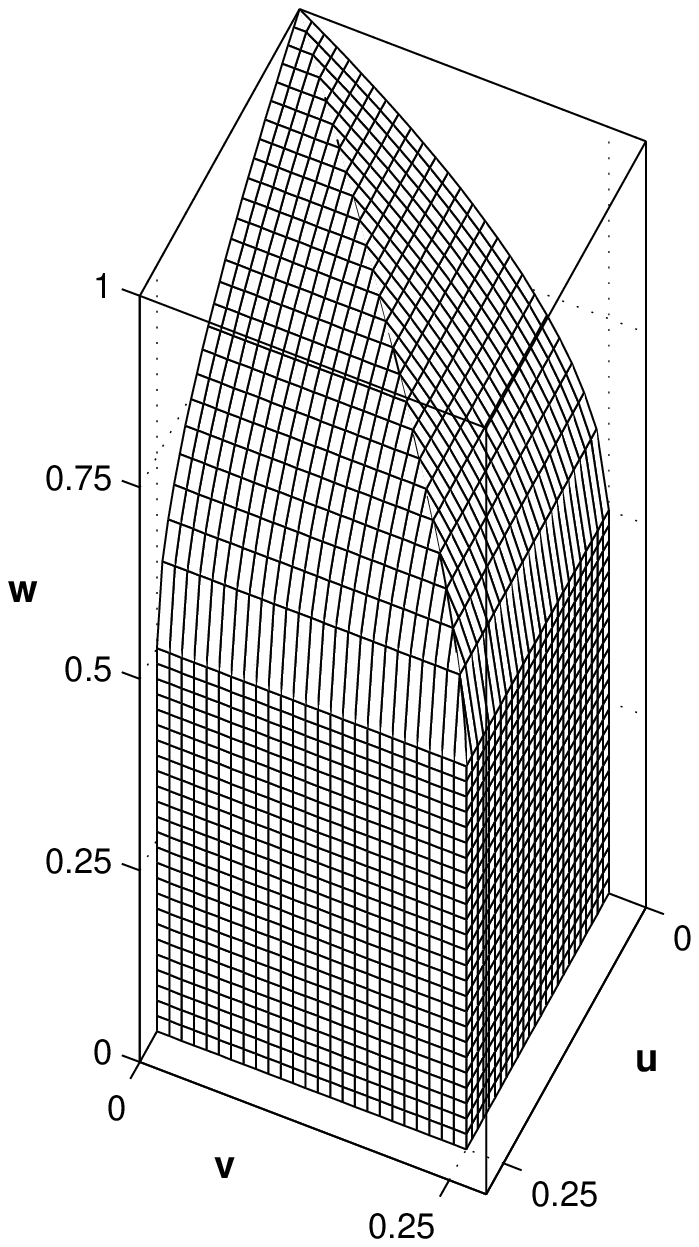} \\ [0.4cm]
\mbox{\bf (a)} & \mbox{\bf (b)}
\end{array}$
\end{center}
\caption{\footnotesize \mbox{\bf (a):} The critical
line, combined from $u=1/4$, $w=\lambda_+(u)$, together with plots of the zeros
of $|P_{N,\{q\} }|$ for random sequences $\{q\}$ and for sequences $AB^{N-1}$
and $A^{\frac{N}{2}}B^{\frac{N}{2}}$
at different values of $N$, in the $u=v$ plane.
\mbox{\bf (b):} The plot of the critical surface in the full parameter space of the
$( u,v,w)$.}
\label{zeros}
\end{figure}

For both $u$ and $w$ different from zero the general expression for
(\ref{AJ^M}) becomes highly non-trivial.
To get a better idea of the distribution of zeros, we used the program
{\sc Matlab} to plot some of the roots of $|P_N(u,w)|$.
This is in principle straightforward, but for increasing $N$ requires
considerable computing power, because the number
of sequences grows like $2^N$.
A superposition of several such plots is shown in Fig.~\ref{zeros}a.\footnote{Typical
$N$-values for the sequences ranged between 50 and 200,
and the total number of sequences analyzed was on the order of 50.}
From this picture it appears that the singularity-free region for $u<1/4$ found
for $w=0$ is also present for $w>0$, up to some value of about 0.5,
and then starts to shrink toward zero. Because of the limit on $N$ in our numerical
study, we could not determine whether the value $u=0$ is reached
at $w=1$ or rather some small $u=\epsilon >0$.

What we can do analytically to put more stringent bounds on the envelope
is to study specific sequences and find their
locus of zeros asymptotically for $N\rightarrow\infty$.
We find that $(u,w)=(\frac{1}{4},w)$ continues to be an
accumulation point for the zeros of
the sequences $A B^{N-1}$ for all values of $w\leq 1$.
For sequences of the form
$A^{\frac{N}{2}}B^{\frac{N}{2}}$ it can easily be shown analytically that
a solution of $Tr(A^{\frac{N}{2}}B^{\frac{N}{2}})=0$ for
$N\rightarrow\infty$ is given by $w=\lambda_+(u)=(1+\sqrt{1-4u})/2$.
For the alternating sequences $(AB)^{N/2}$ we find instead a locus
given by $uw=4/27$, which is tangent to the curve
$w=\lambda_+(u)$ in the point $(u,w)=(2/9,2/3)$, and otherwise lies completely to
the right of it (in a $u$-$w$-plot like that of Fig.\ \ref{zeros}a).
Many other sequences with regular distribution patterns for $A$- and
$B$-matrices whose large-$N$ limit we have studied seem to have an
asymptotic locus of zeros which is a smooth curve lying in between
$w=\lambda_+(u)$ and $w=4/(27 u)$, and passing through their common
point $(2/9,2/3)$.

Based on these findings, and in the absence of any evidence to the contrary,
we conjecture that the $(u,w)$-region free of singularities is defined by the
simultaneous conditions
$u< 1/4$ and $w<\lambda_+(u)$, whose boundary is given by the red
curve in Fig.\ \ref{zeros}a. To its right we have plotted the locations of
zeros, both of random sequences corresponding to some $q_i$'s, and of
special sequences we have been able to treat exactly.
Interestingly, this singular curve can be characterized in terms of the eigenvalues
$(\lambda_+,\;\lambda_-,\;w)$ of the matrix $A$, as the curve along which
the largest two of the eigenvalues are degenerate. For $u>1/4$, the $\lambda_\pm$
are distinct and complex, and thus cannot be equal to $w$ (which is real).
For $u=1/4$, $\lambda_\pm$ are both equal to $1/2$ and larger than $w$,
as long as $w<1/2$. For $w\geq 1/2$, the two largest eigenvalues are
exactly defined by $w$ being equal to the larger one of the pair $\lambda_{\pm}$.
This suggests the existence of an analytical argument for why
such a degeneracy gives rise to roots of $|P_N|$, but we have not been
able to find it.

Following this line of argument further, we can give yet another
characterization of the singular line.
It can be checked that the trace $|P_N|$ is always real.
Denoting the eigenvalues of $P_N$ by $\lambda_i$,  $i=1,2,3$,
their sum must therefore be real. Their product is the determinant of $P_N$,
which from (\ref{AJ^M}) is easily computed as the product of the determinants,
$\lambda_1\lambda_2\lambda_3=\lambda_+^N\lambda_-^Nw^N=(uw)^N$,
which is always real and positive for $u,w>0$.
This leaves only three possibilities for the signature/character of the $\lambda_i$:
$(+++)$, $(+--)$ or $(+\;c\;\bar{c})$, where $c$ and $\bar{c}$ denotes complex
conjugates.
Of course with signature $(+++)$ the trace can never be zero, and this is
precisely the
case in the region $u<1/4$ and $w<\lambda_+(u)$. From this we cannot
go directly to
signature $(+--)$ because two of the eigenvalues would have to pass
through zero, in which case the determinant would become zero,
leading to a contradiction. The transition occurring at the singular line
should be to the region with $(+\;c\;\bar{c})$. From there, once two of the eigenvalues
are complex, their real part can become negative and also the trace can
become zero. This scenario is confirmed by numerical computations, but
a general, algebraic proof is at this stage still missing.

\subsection{u $\neq$ v}

Using the same mixture of analytical and numerical methods as for the case $u=v$,
and with a comparable computational effort\footnote{Amongst other things, by
systematically ``scanning" the three-dimensional coupling space in $(v,w)$-planes
for various fixed values of $u$.}
we have found a similar picture in the full three-dimensional parameter space with
$u\neq v$. The critical surface is determined by the same relations as for
$u=v$ above, but with $u$ substituted by $\max (u,v)$. That is, for $u<v$ the
critical surface is given by $v=1/4$ for $w\leq 1/2$
and $w=\lambda_+(v)$ for $w>1/2$, while for $u>v$ it is
given by $u=1/4$ for $w\leq 1/2$
and $w=\lambda_+(u)$ for $w>1/2$.
Since our extensive numerical and analytical checks have turned up no
contradictions to this picture, we will in the
following assume it to be correct. The critical surface is depicted in
Fig.\ \ref{zeros}b.

\section{The large-$N$ limit} \label{sec-replica}

We will now assume that we are inside the region free of singularities. In this region
the partition sum is convergent and its large-$N$ limit well defined.
The boundary of this region is the critical line (or critical surface for $u\neq v$).
We want to characterize the (non-)analytic character of the partition function
$Z_N(u,v,w)$ on this boundary
{\it after} the limit for large $N$ has been taken (see the discussion in the previous section).
Initially the evaluation of this function seems an insurmountable task, because
according to (\ref{Zinv}) and (\ref{ZZZ}) we have to sum over the {\it inverses}
of the matrix traces resulting from the application of the inversion formula.
However, it was exactly with this difficulty in mind that we introduced
the reformulation of the partition function in terms of random matrix products
in Sec.\ \ref{sec-matrices} above. Techniques available for such products of
random matrices will help us to estimate precisely the limit we are interested in,
namely
\beq
L_{-1}(u,v,w)  =
\lim_{N\rightarrow \infty} \frac{1}{N}\ln \left\langle\frac{1}{|P_N(u,v,w)|}\right\rangle
= -\ln 2 +\lim_{N\rightarrow \infty} \frac{1}{N}\ln Z_N(u,v,w,\Delta t=1),
\eeq
where, as in (\ref{Z}) above, the average has been taken over all random
sequences $\{q_j\}$.
(The reason for the notation $L_{-1}$ will become clear soon.)
The aim of this section is to compute  $L_{-1}(u,v,w)$, at least in a perturbative expansion
around the critical point (which still needs to be identified).

\subsection{Generalized Lyapunov exponents}

\begin{figure}[h]
\centering
\includegraphics[width=12cm]{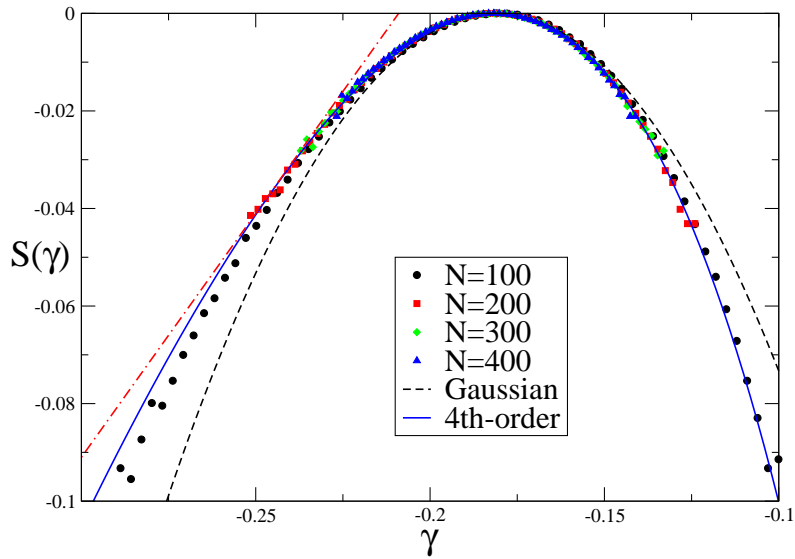}
\caption{Determining the function $S(\g)$ numerically for $u=v=0.24$ and $w=0.1$.
The Gaussian approximation
is shown as a dashed line. The blue continuous line is a fourth-order fit. The straight
(dashed-dotted) line
on the left has
slope 1, and is tangent to $S(\g)$ in $\g^*_{-1}$. In this case, the Gaussian approximation
is not sufficient to compute $L_{-1}$ from (\ref{lns}),
because in the point $\g = \g^*_{-1}$, $S(\g)$ is very different
from the Gaussian fit. The figure illustrates a ``worst-case scenario" in that for
the chosen parameter values the deviation of the actual curve from the Gaussian
approximation is maximal, see also Fig.\ \ref{replichefig}.
See appendix A-III for details on the numerical method.}
\label{Sgamma}
\end{figure}

For a given sequence $\{q_j\}$, define a real number $\gamma$ by
\beq
\gamma=\frac{1}{N}\ln |P_N|  ,
\eeq
where $|P_{N}|$ is a shorthand notation for $|P_{N,\{q\}}(u,v,w)|$.
Since $\{q_j\}$ is random, $\gamma$ will be a random variable as well;
its probability distribution has, for large $N$, the general form~\cite{crisanti}
\beq
\pi_N(\gamma)\propto e^{N S(\gamma)} .
\eeq
The function $S(\g)$ is called {\it large deviations function}; it is a convex
function that has a maximum in $\g=\bar\g$, where $\bar\g$ is defined as
\beq\label{barg}
\bar\g=\lim_{N\rightarrow\infty}\frac{1}{N}\ln |P_N|
\eeq
and called the \emph{maximum Lyapunov characteristic exponent}.\footnote{In general,
the Furstenberg theorem
\cite{crisanti} guarantees that $\bar\g$
exists with probability 1 and is a non-random quantity,
$i.e.$ $\bar\g=\lim_{N\rightarrow\infty}\frac{1}{N}\langle\ln |P_N|\rangle$.
This means that $\gamma$ is a \emph{self-averaging} quantity, and that
$S(\gamma)$ must be a function peaked around $\bar\g$.
Note that instead $|P_N|=e^{N\gamma}$, or an integer power of $|P_N|$ (including $|P_N|^{-1}$),
is not in general a
self-averaging quantity: it will be shown later in this
section that
$\la |P_N| \ra = e^{N L_1}$ and $\la |P_N|^2 \ra = e^{N L_2}$ where the $L_n$ are $O(1)$, so that,
as long as $L_2 > 2 L_1$,
$\langle (|P_N|-\langle |P_N| \rangle)^2 \rangle / \langle |P_N|\rangle^2 \sim e^{N(L_2 - 2 L_1)} \gg 1$
.}
We will choose $S(\bar\g)=0$, so that the distribution $\pi_N(\g)$ is
normalized at the leading order for $N\to \io$:
\beq
\int d\g \, e^{N S(\g)} \sim e^{N S(\bar\g)} = 1 .
\eeq
We can then write
\beq
  \left\langle\frac{1}{|P_N(u,v,w)|}\right\rangle=\langle e^{-N\gamma}\rangle =
  \int d\gamma \, e^{N[S(\gamma)-\gamma]}\sim e^{N[S(\gamma^{\ast})-\gamma^{\ast}]}  ,
\eeq
where $\gamma^{\ast}$ is the solution to the saddle point equation\footnote{We are assuming that $S(\g)$ is
analytic and convex, so that the solution to the saddle point equation exists and is unique. Even if we cannot prove
this assumption, it is strongly supported by numerical simulations, see Fig.~\ref{Sgamma}. The same assumption guarantees
that $L_n$ is an analytic function of $n$ and is at the basis of the replica method used in
the next subsection.}
\beq
  \frac{\partial S(\gamma)}{\partial\gamma}=1  .
\eeq
Similarly, one can define the Legendre transform of $S(\g)$ (or {\it generalized
Lyapunov exponent}) by
\beq
\label{lns}
L_n = \lim_{N \to \io} \frac{1}{N} \ln \langle |P_N|^n \rangle =
\lim_{N \to \io} \frac{1}{N} \ln \langle e^{n\g N} \rangle = S(\g^*_n) + n\g^*_n ,
\eeq
where $\g^*_n$ is the solution of
\beq
\label{gn}
\frac{dS}{d\g}=- n .
\eeq
Clearly, for integer $n$, $e^{N L_n}=\la |P_N|^n \ra$ is the $n$-th moment of $|P_N|$.
If one approximates $S(\gamma)$ by a
Gaussian\footnote{A zeroth-order approximation, corresponding to $\s=0$,
is $\pi_N(\gamma)=\delta(\g-\bar\g)$,
which gives the trivial result $L_n=n \bar\g$.}, $S(\g)=-\frac{(\g-\bar\g)^2}{2\s^2}$, one easily finds
\beq
\label{gauss}
\gamma^{\ast}_n=\bar{\gamma} + n \sigma^2  ,
\hskip30pt
L_n = n \bar\g + \frac{1}{2} n^2 \s^2  .
\eeq
In this case the knowledge of the first two moments $L_1$ and $L_2$ is
equivalent to the knowledge of
$\bar \g$ and $\s$, and determines the full curve $L_n$, now regarded
as an analytic function for all values of $n$.
If $S(\g)$ is not exactly Gaussian, a systematic expansion of $L_n$ in powers of $n$,
\beq\label{serieL}
L_n = \sum_{k=1}^\io \frac{l_k}{k!} n^k ,
\eeq
can be obtained starting from the expansion of $S(\g)$ in powers
of $\g - \bar\g$ and solving eq.~(\ref{gn})
order by order, given that $n=O(\g - \bar\g)$.
The knowledge of the first $k$ integer moments $L_k$ is equivalent to the
knowledge of the first $k$ derivatives of $S(\g)$ in $\g=\bar\g$ and
allows us to reconstruct $L_n$ up to a given order of approximation.
Therefore, one possible way of investigation is to measure $S(\g)$ numerically,
use a Gaussian
fit to estimate $\bar{\gamma}$ and $\sigma^2$, and eventually compute the
corrections coming from the cubic, quartic, $\ldots$ terms of $S(\g)$. An illustrative numerical result
for the function $S(\g)$ is plotted in Fig.~\ref{Sgamma} for $u=v=0.24$, $w=0.1$,
where the deviation from Gaussianity is maximal. The Gaussian approximation becomes
better as the critical point is approached. The insights gleaned from this numerical
analysis become most powerful when combined with a different tool for computing
the moments $L_n$, based on the so-called replica trick, to which we will turn in
the next subsection.

\subsection{Replica trick}\label{replica-sec}

A more efficient strategy is to compute the integer moments $L_n$ directly
with the help of the replica trick.
For $n$ a positive integer and for any sequence
of matrices $M_j$, $j=1,\cdots,N$, it is easy to show that
\beq
\left(\Tr\prod_j M_{q_j} \right)^n = \Tr \prod_j M_{q_j}^{\otimes n}  ,
\eeq
where $\otimes$ is the tensor product. Then, if the $M_j$ are independently
and identically distributed,
\beq
\langle |P_N|^n\rangle=\left\langle \left(\Tr\prod_jM_{q_j} \right)^n\right\rangle
=\Tr\left\langle \prod_j M_{q_j}^{\otimes n}\right\rangle =
\Tr \prod_j \la M_{q_j}^{\otimes n} \ra = \Tr \la M^{\otimes n} \ra^N \sim \nu_n^N .
\eeq
where $\nu_n$ is the largest eigenvalue of the matrix $\la M^{\otimes n} \ra$, which can
be easily evaluated for $n$ small.
Thus one has, for positive integer $n$,
\beq
L_n = \ln \nu_n .
\eeq
Knowing the function $L_n$ for integers $0 < n\leq k$ allows one to compute the first
$k$ coefficients $l_1,\cdots,l_k$ of the $n$-expansion (\ref{serieL}) simply by
solving a linear system.
This yields an approximate expression for
$\bar\g = \lim_{n\to 0} \frac{d L_n}{d n} = l_1$ and for
$L_{-1}$, which is the quantity we want to compute.
Obviously the method assumes that $L_n$ is an analytic function of $n$ and
would break down if there were a singularity at some value of $n$.

\begin{figure}[t]
\centering
\includegraphics[width=10cm]{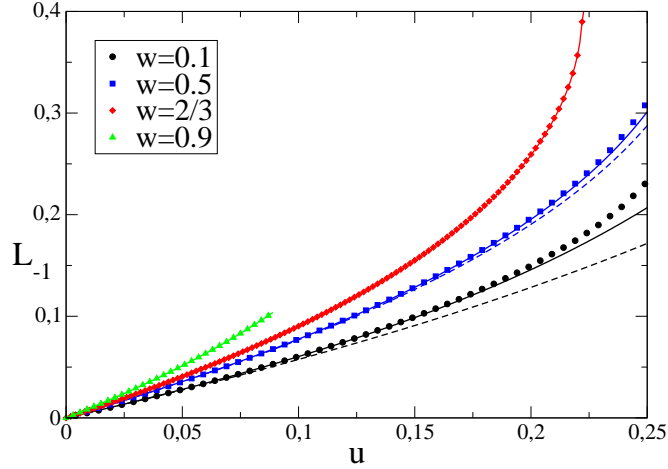}
\caption{
The numerical result for $L_{-1}$ as a function of $u(=v)$ for $w=0.1,0.5,w_c,0.9$.
For $w \geq w_c=2/3$, the simplest approximation $L_{-1}=-L_1$ (dashed line) works well.
For $w=0.1$ and $0.5$,
the Gaussian approximation, $L_{-1} = L_2 - 3 L_1$ (continuous line), is needed to approximate
the data better. For small $w$ and $u\to 1/4$, it is clear that additional
higher-order corrections are necessary, see also Fig.~\ref{Sgamma}. However, the numerical
data do not indicate a divergence of the derivative of $L_{-1}$ with respect to $u$
for $w \neq w_c$.
}
\label{replichefig}
\end{figure}

\subsubsection{First moment}

The simplest approximation is $L_n = l_1 n$. The coefficient $l_1$ is then
equal to $L_1$ and is simply the logarithm of the largest eigenvalue of the matrix
$\langle M \rangle = (A+B)/2$, that is,
\beq\begin{split}
&\nu_1(u,v,w) = \frac{1}{4} \left( 2 + w + \sqrt{(2-w)^2 - 4(u+v)} \right) , \\
&L_1(u,v,w) = \ln \n_1(u,v,w)  .
\end{split}\eeq
This approximation is equivalent to neglecting the fluctuations of $\g$ since
\beq
e^{NL_n} = \left\langle |P_N|^n \right\rangle = e^{nNL_1}= \la |P_N|\ra ^n   .
\eeq
The function $\nu_1(u,v,w)$ has a singularity at
\beq
u+v=\left(\frac{2-w}{2}\right)^2,
\eeq
which intersects the critical surface only at the {\it critical point} $(u_c,v_c,w_c)
\equiv ( 2/9,2/9,2/3)$.
This point is therefore a good candidate for a singularity
of $L_{-1}$, if it does not cancel at the next order.

\subsubsection{Second moment: Gaussian approximation}

Including the first correction is equivalent to the Gaussian approximation and is obtained
by considering $L_n = l_1  n + \frac{1}{2} l_2 n^2$ as in (\ref{gauss}) above. One has
\beq
l_1 = \bar\g = 2 L_1 - \frac{1}{2} L_2 , \hskip.5cm
l_2 = \s^2 = L_2 - 2 L_1  ,
\eeq
with $L_1=\ln \n_1$ as before and $L_2=\ln \n_2$, the logarithm of the
largest eigenvalue of the $9\times 9$ matrix
$\langle M^{\otimes 2} \rangle = (A^{\otimes 2}+B^{\otimes 2})/2$. For our searched-for
expression this implies $L_{-1}=L_2 - 3L_1$.
We have calculated the eigenvalue $\n_2$ with the help of {\sc Mathematica} for
$u=v$, but refrain from reporting it here, because it is long and not particularly
illuminating.
However, the expressions simplify considerably on the critical line $u=v=w-w^2$,
where one has
\beq\label{nue2}
\n_2(w-w^2,w) =
\begin{cases} &w^2  \hskip206pt  w > \frac{2}{3}  , \\
&\frac{1}{4}\left[2-2w+w^2 + \sqrt{4(1-w)^2 + w^2 (2-3w)^2}\right] \hskip15pt w <\frac{2}{3}  .
\end{cases}
\eeq
What is rather remarkable about this result is that for $w>2/3$ we have
$\n_2 = \n_1^2$, \ie $L_2 = 2L_1$.
It means
that the approximation $L_n=nL_1$ is {\it exact}\footnote{To prove this statement one should
prove that $\n_n = \n_1^n$ for all $n$. Although this ought to be possible, we have
limited ourselves here to checking, by sampling random values on the critical
line above $w>2/3$, that it is
true for $n$ large but finite.} on the critical line for $w>2/3$.
On this line we then obtain $L_{-1} = -\ln \n_1 = -\ln w$.

More generally, we can write $L_{-1}=-L_1 + (L_2 - 2L_1) = -L_1 + \D L_2$ and
study the behaviour of $\D L_2$ close
to the critical line.
For later use in our scaling analysis, we
consider two linearly independent types of perturbation away from a
given point on the critical line, labelled by its $w$-coordinate $w_0$.
The first one is for $w = w_0 - \d$
and $u=v=w-w^2$, thus describing motion
along the critical line, for which we find
\beq\label{deltacase}
\D L_2 = \begin{cases}
0 \hskip98pt \text{ for } w_0 > 2/3,\\
\kappa_1 \delta^2+O(\delta^3) \hskip45pt \text{ for } w_0 = 2/3,\\
\kappa_2+\kappa_3\delta +O(\delta^2) \hskip27pt \text{ for } w_0 < 2/3.
\end{cases}
\eeq
The second one moves away from the critical line inside the diagonal
plane $u=v$, and toward the interior of the singularity-free region
according to $u = w_0-w_0^2-\ee$, yielding
\beq\label{epscase}
\D L_2 = \begin{cases}
\kappa_4 \ee^4 + O(\ee^5) \hskip62pt \text{ for } w_0 > 2/3  , \\
\kappa_5 \ee + \kappa_6 \ee^{3/2} + O(\ee^2) \hskip29pt \text{ for } w_0 = 2/3  , \\
\kappa_7 + \kappa_8 \ee + O(\ee^2) \hskip45pt \text{ for } w_0 < 2/3
       \end{cases}
\eeq
where the $\kappa_i$ in (\ref{deltacase}) and (\ref{epscase}) are numerical constants.
Interestingly, these corrections
do not change the divergent term in the first derivatives of
$L_{-1}$ with respect to $u$ and $v$ at the critical point, which govern
the presence or otherwise of an infinite-volume limit. Neither, since $\nu_2$
according to (\ref{nue2}) has no further singularities, can singularities appear at
second order at points with $w_0\not= 2/3$.
In summary, since all correction terms give only finite contributions to the
derivatives, it is clear that our first-order analysis already correctly identified
$(u_c,v_c,w_c)= ( 2/9,2/9,2/3)$ as the only point at which an
infinite-volume limit exists. This is confirmed by the numerical results for
$L_{-1}(u,v=u,w)$
reported in Fig.~\ref{replichefig}.

Since $\D L_2$ is proportional to $\ee$ at the critical
point, we expect it to contribute to the calculation of
the Hamiltonian. This would at first seem to necessitate an analytic determination for
the eigenvalues of $(A^{\otimes 2}+B^{\otimes 2})/2$ for  $u\neq v$,
which is currently out of reach. Fortunately,
all that is needed is a perturbative evaluation after inserting an ansatz for
the scaling of $u$, $v$ and $w$ near
the critical point, which is a perfectly feasible task we will perform in Sec. 9 below.

\subsubsection{Higher moments: beyond the Gaussian approximation}

The precision in determining $L_{-1}$ can be improved further by
computing $L_3$, $L_4$, etc. In Fig.~\ref{replichefig} we
report our numerical findings for $L_{-1}$ (see appendix \ref{App-numeric}),
together with the first- and second-moment approximation just described.
For $w>2/3$ the first-order approximation is excellent, and the numerical
deviations from $\D L_2 \sim 0$
for $w>2/3$ and any value of $u$ are very small.
Around $w=1/2$, the Gaussian correction starts to
be observable. For $w<1/2$ and $u \sim 1/4$, the correction is large, c.f.
Fig.~\ref{Sgamma}, and higher-order corrections have to be taken into account
to correctly reproduce the numerical result.

For approaches to the critical point, we have found that $\D L_r$, the incremental correction
for the calculation of $L_{-1}$ which results from adding the $r$'th one to
that of the first $r-1$ moments, scales like $\ee^{r/2}$ and thus
is not expected to contribute to the calculation of the Hamiltonian. We have confirmed
this by perturbative calculations for $L_3$.

\section{Continuum limit, canonical scaling and properties of the slices} \label{sec-continuum}

As usual in dynamically triangulated models of quantum gravity, our next step will be to search
for a continuum limit in which the details of the discretization procedure will be ``washed out",
and only universal, physical properties will remain. A necessary part of this limit is to take our
short-distance cutoff, the edge length $a$, to zero, while keeping the {\it physical} spacetime
volume $\mathcal V$ finite. This is only possible if the number of building blocks
in the simplicial manifold (the {\it discrete} spacetime volume $V$) simultaneously goes to
infinity. More precisely, we have to take the continuum limit
following a trajectory (parameterized by $a$) in the coupling-constant space which ends up
(for $a=0$) in a point
where the expectation value of the number of building blocks diverges, and which approaches this point
in such a way that the physical volume stays finite. In other words, we have to renormalize the bare
cosmological (and possibly other coupling) constants so that relevant
physical quantities remain finite.

Our first step, the identification of a critical point with suitable properties, follows from the
discussion of the previous section: the only point where the derivatives of the partition
function\footnote{Remember that the average volume
is given by the derivative of the logarithm of the partition function with respect to the bare cosmological
constant, which translates into a linear combination of $u\frac{\partial}{\partial u}$,
$v\frac{\partial}{\partial v}$ and $w\frac{\partial}{\partial w}$, see (\ref{sandwich-volume}).}
diverge is $(u_c,v_c,w_c)= ( 2/9,2/9,2/3)$.
As a second step we will assume a canonical scaling, where the critical point is approached according
to the canonical dimensions of the corresponding continuum coupling constants. To lowest
non-trivial order in $a$, these are by definition
\beq \label{canonical}
k\simeq k_c+\frac{a}{G}, \hspace{.3cm} \ln x\simeq -\l_{b,c}-a^2 X, \hspace{.3cm}
\ln y\simeq -\l_{b,c}-a^2 Y,
 \hspace{.3cm} \l\simeq \l_c+a^3\L ,
\eeq
for the inverse Newton, the two boundary cosmological and the bulk cosmological constants.
In addition, a canonical scaling would require $T=a t$ and ${\cal N}=aN$ for the time and
spatial extensions.

Translated into the variables $u$, $v$ and $w$ (which were defined in (\ref{laplace})),
the canonical scaling becomes
\beq \label{canonical2}
 u=\frac{2}{9}\ e^{2 a c_1/G-2a^2 X-2 a^3 b_1\L}, \hspace{.3cm}
 v=\frac{2}{9}\ e^{2 a c_1/G-2a^2 Y-2 a^3 b_1\L}, \hspace{.3cm}
 w=\frac{2}{3}\ e^{-a c_2/G-a^3 b_2\L} .
\eeq
Already without detailed inspection we can anticipate difficulties with the standard
interpretation of the bulk and boundary cosmological constants as the couplings
conjugate to the bulk volume and boundary areas.
Since $k$ scales with the lowest power of $a$ when approaching the critical point,
terms proportional to $1/G$ will be the ones which survive in the lowest-order expressions
of the continuum limit,
unless unexpected cancellations occur.

We can check this immediately by looking at what this particular choice of scaling
implies for geometrical quantitites like the volume of sandwich geometries.
As argued in the previous section, the second moment in the replica method only contributes to the
next-to-leading order, which therefore we will need for the Hamiltonian,
but not to recover the continuum expression for the volume.
This means that as a starting point for our computation we can take
the simple expression
\beq
Z_N\sim e^{-N\ln \n_1(u,v,w)}
\eeq
as the partition function of a sandwich geometry. Assuming canonical scaling for
the three-volume and using previous definitions, we have
\beq \label{sandwich-volume}
\begin{split}
\langle {\mathcal V} \rangle &= \lim_{a\rightarrow 0} a^3 \langle V \rangle,\\
\langle V \rangle &= \left ( -\frac{\partial}{\partial \l} \ln Z_N\right )=
\left ( 2 b_1 u\frac{\partial}{\partial u}+ 2 b_1 v\frac{\partial}{\partial v} +
 b_2 w\frac{\partial}{\partial w} \right ) \ln Z_N ,
\end{split}
\eeq
which after inserting the scalings (\ref{canonical2}) and taking the limit $a\rightarrow 0$ yields
\beq
\langle {\mathcal V} \rangle = a^{5/2} N \frac{2b_1+b_2}{4\sqrt{(c_2-2c_1)/G}}
\eeq
to lowest order in $a$.
As expected, only the renormalized Newton constant $G$ appears in the continuum expression
for the volume. In addition, this expression
indicates an anomalous scaling of either the time variable $t$ or the variable $N$ measuring the
linear spatial extension. For example, if we insist on the canonical scaling
${\cal N} = a N$ for the finite continuum counterpart of $N$, we find that the volume
goes to zero like $a^{3/2}$ instead of $a$ as would be required for a canonical scaling $T=at$
of the time variable.

If instead we set $k=0$ identically, corresponding to working with a bare action which only
contains a cosmological term
the sandwich volume scales canonically according to
\beq
\langle {\mathcal V} \rangle = a^2 N \frac{2b_1+b_2}{4\sqrt{X+Y}}=a {\cal N }\frac{2b_1+b_2}{4\sqrt{X+Y}}.
\eeq

Since we do not want to touch the canonical scaling of the three-volume, which would affect
the interpretation of the model in a fundamental way, one possible solution is
to choose the constants $c_i$ such that $c_2=2 c_1$ and the
order-$a$ terms cancel each other.
In this case we would obtain
\beq
\langle {\mathcal V} \rangle = a^2 N \frac{2b_1+b_2}{4\sqrt{X+Y-\frac{3}{4}c_2^2 /G^2}} .
\eeq
The scaling now {\it is} canonical, and the sandwich volume is governed by both the
boundary cosmological constants and the Newton constant. We will see below that
related issues appear when we try to derive the continuum Hamiltonian.

Another quantity one can consider is the total, integrated scalar curvature
${\cal R}_{tot}\equiv\int d^3 x \sqrt{g}\ {\cal R}$.
Its counterpart at the discrete level is exactly the term $R_{tot}$
multiplying $k$ in the gravitational action, for which we have used the standard
Regge prescription. We can express the average total curvature
as a derivative with respect to $k$ of the partition function, and then take the continuum limit.
The analogues of relations (\ref{sandwich-volume}) are given by
\beq
\begin{split}
\langle {\cal R}_{tot} \rangle &= \lim_{a\rightarrow 0} a\langle R_{tot}\rangle,\\
\langle R_{tot}\rangle &= \frac{\partial \ln Z_N}{\partial k}=
 -c_3 N+\left( c_1\left( 2u\frac{\partial}{\partial u}+ 2v\frac{\partial}{\partial v}\right)
 -c_2w\frac{\partial}{\partial w}\right)\ln Z_N .
\end{split}
\eeq

Since our spacetime sandwiches in the continuum limit have only infinitesimal
thickness and thus infinitesimal volume, it is more appropriate to work with
the average curvature per volume,
\beq
\bar{\cal R} \equiv \frac{\langle {\cal R}_{tot} \rangle }{\langle {\cal V}\rangle } .
\eeq
It is a well-known feature of dynamically triangulated models of gravity that
this quantity generically diverges like $a^{-2}$, unless specific
cancellations occur.
We find here not only the same behaviour, but even the same algebraic expression
to leading order, namely,
\beq
\bar{\cal R} =  \frac{1}{a^2}\left(\frac{2c_1-c_2}{2b_1+b_2}\right),
\eeq
regardless of
whether we assume the scaling $k=0$, or the canonical $k\simeq k_c+a/G$.
However, in line with our earlier remarks, {\it if} one makes the choice\footnote{
This can be achieved by fixing the finite relative factor $r$ between the
time- and space-like (squared) edge lengths of the triangulations (c.f. appendix A-I).
Using eq.\ (\ref{alphabeta2}), it corresponds to taking $r\simeq 0.724$.} $2 c_1=c_2$,
a different, less divergent, behaviour is found.
Combining this condition and canonical scaling, we obtain
\beq
\label{rav1}
\bar{\cal R} =  \frac{1}{a}\left(\frac{3c_2^2/G}{4b_1+2b_2}
+\left( \frac{c_2-4c_3}{2b_1+b_2} \right)\sqrt{X+Y-\frac{3}{4}c_2^2 /G^2}\
\right) .
\eeq
For $k=0$ we find instead
\beq
\label{rav2}
\bar{\cal R} =  \frac{1}{a} (c_2-4c_3)\frac{\sqrt{X+Y}}{2b_1+b_2}
-\frac{3}{2}\left(\frac{c_2}{2b_1+b_2}\right)(X+Y).
\eeq
We will encounter somewhat similar scaling relations in the discussion of
the quantum Hamiltonian in Sec.~\ref{sec-hamiltonian} below.

\section{Transfer matrix for areas}\label{sec-gluing}

One way to construct the propagator for finite times $t$ of a statistical model is
by iteration of the transfer matrix $\hat T$, which in our case would take the
form  $\langle g_2|\hat T^t|g_1\rangle$. Although the determination of the complete
transfer matrix for our three-dimensional model is out of reach, we have already
argued in Sec.\ \ref{sec-part} that most of the detailed dependence of the propagator
on the boundary geometries $g_i$ will be dynamically irrelevant, because
of the absence of local degrees of freedom in the three-dimensional gravity theory.
Following \cite{ABAB1}, keeping track only of the boundary areas $A_i$ in the one-step propagator
as we have been doing throughout this work, should be enough\footnote{Possibly up to an additional
dependence on global Teichm\"uller
parameters.}  to determine the finite-time
behaviour of the propagator in the continuum limit.
Introducing the ``area states"
\beq
|A\rangle = \frac{1}{\sqrt{\NN(A)}}\sum_{g_{|A}}|g_{|A}\rangle,
\eeq
where $\NN(A)$ is the number of triangulations of a given area $A$ and $g_{|A}$ any triangulation
with a given total number of triangles $A$, we obtain that
\beq \label{Z-Sqrt}
\begin{split}
Z(x,y,\Delta t=1)&=\sum_{A_1,A_2}x^{A_1}y^{A_2}\sum_{g_{|A_1},g_{|A_2}}\langle g_{|A_2}|\hat T|g_{|A_1}\rangle=\\
&=\sum_{A_1,A_2}x^{A_1}y^{A_2}\langle A_2|\hat T|A_1\rangle\sqrt{\NN(A_1)\NN(A_2)}.
\end{split}
\eeq
Assuming now in line with our earlier reasoning (see also \cite{ABAB1}) that
\beq \label{conjecture}
\langle A_2|\hat T|g_{|A_1}\rangle-\langle A_2|\hat T|g'_{|A_1}\rangle\rightarrow 0 \hskip29pt \text{for}
\hskip15pt A_2,A_1\rightarrow\infty,
\eeq
for all pairs $g$, $g'$ of boundary geometries of the same area, the completeness relation
\beq
\int dA |A\rangle\langle A|=\mathbb{I}
\eeq
holds in the large-area limit and we can use $\langle A_2|\hat T|A_1\rangle$
as our transfer matrix.
One remaining problem is the appearance of the square-root term $\sqrt{\NN(A_t)\NN(A_{t+1})}$
in (\ref{Z-Sqrt}), which
we will have to deal with either before or after performing the inverse Laplace transform of $Z(x,y,\Delta t=1)$.
Given that the number of triangulations of a given area $A$ scales like
\beq \label{entropy-A}
\NN(A)\sim A^{-\a} e^{\l_0 A},
\eeq
with $\a =1/2$ as we will show at the end of this section,
one possibility is to apply to  $Z(x,y,\Delta t=1)$ a fractional derivative
operator
\beq
\sum_{A_1,A_2}x^{A_1}y^{A_2}\langle A_2|\hat T|A_1\rangle=\left(\frac{\partial}{\partial\ln x}\right)^{\a /2}
\left(\frac{\partial}{\partial\ln y}\right)^{\a /2} Z(x,y,\Delta t=1)\ .
\eeq
There are several definitions of fractional derivatives in the literature; what we need here is an operator
$D^{\a}_x$ such that $D^{\a}_x e^{A x}=A^{\a} e^{A x}$, so that applying $D^{\a/2}_{\ln x}D^{\a/2}_{\ln y}$
term by term to (\ref{Z-Sqrt}) we get rid of the entropy factors (the remaining exponential term
$e^{\l_0 A}$ is unproblematic since it only shifts the location of the critical point).
We need yet another property for our fractional derivative, namely, a ``chain rule", since the
final expression for $Z(x,y,\Delta t=1)$ will involve some function of $x$ and $y$.
In particular, since the final expression is expanded in powers of $\ln x$, it would be nice if
the derivative would act on powers with the simple rule $D^{\a}_x x^n\propto x^{n-\a}$.
An explicit representation of a fractional derivative operator with the desired properties exists
and is reviewed in appendix \ref{App-fracder}.

Unfortunately, at the relevant value $\a=1/2$, for some terms in the final expression for
$Z(x,y,\Delta t=1)$ the fractional derivatives suffer from convergence problems.
These can be circumvented in a somewhat {\it ad-hoc} fashion by integrating by parts or
by introducing a regularization in the fractional integrals,
which should be removed only after the ordinary derivative has been applied.
(As explained in appendix \ref{App-fracder},
the fractional derivative is actually defined as an ordinary derivative acting on a fractional integral.)

A cleaner alternative to get the correct Hamiltonian is to first work out the inverse
Laplace transform and then identify the contribution coming from the entropy factor
and remove it.
This method on the other hand requires some special care for the following reason.
Starting from (\ref{Z-Sqrt}), keeping only the two lowest orders in $a$ (see (\ref{HfromT})), and doing an inverse
Laplace transform will produce $\NN(A_1)\d(A_1-A_2)$ at the lowest order and some differential operator acting
on $\d(A_1-A_2)$ at the next order. If we divided this operator by $\NN(A_t)$ to obtain the Hamiltonian,
we would make a mistake since at next-to-leading order $A_2\simeq A_1 +a \d A$, which needs to be
taken into account.
To do so let us rewrite the coefficients of the power series (\ref{Z-Sqrt}) as follows:
\beq \label{H-comm}
\begin{split}
\langle A_2|\hat T|A_1\rangle &\sqrt{\NN(A_1)\NN(A_2)} =
\langle A_2|\sqrt{\NN(\hat A)}\ \hat T\ \sqrt{\NN(\hat A)}\ |A_1\rangle = \\
 &=\langle A_2|\NN(\hat A)\hat T|A_1\rangle
 + \langle A_2|\sqrt{\NN(\hat A)}\left[\hat T,\sqrt{\NN(\hat A)}\right] |A_1\rangle =\\
 &=\NN(A_2)\langle A_2|(1-a\hat H)|A_1\rangle
 -a\sqrt{\NN(A_2)}\langle A_2|\left[\hat H,\sqrt{\NN(\hat A)}\right]|A_1\rangle +O(a^2) ,
\end{split}
\eeq
where we have introduced the area operator $\hat A|A_1\rangle=A_1|A_1\rangle$.
If we define the auxiliary Hamiltonian $\hat H'$ by
\beq \label{H-aux}
\langle A_2|\hat T|A_1\rangle \sqrt{\NN(A_1)\NN(A_2)} := \NN(A_2)\langle A_2|(1-a\hat H')|A_1\rangle
 +O(a^2) ,
\eeq
we can invert (\ref{H-aux}) by using (\ref{H-comm}) to find the real Hamiltonian
\beq
\label{H-inversion}
\hat H = \hat H' + \sqrt{\NN(\hat A)} \left[\hat H',\frac{1}{\sqrt{\NN(\hat A)}}\right] .
\eeq
It turns out that also with this method convergence problems appear in some of the
inverse Laplace transforms to be performed. Fortunately, since the divergences
in the fractional derivative method and in the auxiliary Hamiltonian method appear
in different terms, they can be regularized consistently by requiring the end result
in both methods to agree (see appendix \ref{App-fracder}).

We conclude this section by showing that $\a=1/2$ in the scaling relation
(\ref{entropy-A}) for the number of boundary geometries.
The entropy of this class of triangulations can be deduced from its leading critical behaviour.
Summing over all the triangulations with a cosmological weight -- as is appropriate for evaluating
the gravitational partition function in two dimensions -- one has
\beq \label{gamma}
\sum_{\TT} e^{-\l A} \sim \sum_A A^{-\a} e^{-(\l-\l_0)A} \sim (\l-\l_0)^{\a-1} .
\eeq
Of course, the boundaries of our model are nothing but standard (1+1)-dimensional causal
dynamical triangulations with periodically identified time. For precisely this ensemble
we have already established earlier the relation (\ref{Z-2d}), with whose help we will now
be able to find $\a$ analytically. Summing over $t$ to include all triangulations (and
including a factor of $1/2^t$, which is necessary for convergence\footnote{The normalization is needed
in order to take the continuum limit of the partition function at fixed continuum time $T$.
This can be understood
in the following way. If we start with the generating function (\ref{gen2bis}) for the transfer matrix, we find
to leading order {\it twice} the identity operator in the continuum limit. Consequently, if we iterate
(\ref{gen2bis}) $t$ times, we will get a multiplicative factor $2^t$, which has to be removed if we want
to obtain the continuum $T$-propagator.}),
we simply get
\beq \label{2d-sumT}
\sum_t \frac{1}{2^t} Z^{2d}_t = \frac{1}{\ln 2 + \ln \l_+(u)} \sim \frac{1}{\sqrt{1-4 u}},
\eeq
since there is no power-like subleading behaviour in $t$.
Comparing with (\ref{gamma}), we deduce that for the (1+1)-dimensional boundaries we have $\a=1/2$.

\section{The Hamiltonian} \label{sec-hamiltonian}

The continuum dynamics of the model is encoded in the Hamiltonian operator, which
can be derived by taking the continuum limit of the transfer matrix, as we explained before.
Starting from Eq.~(\ref{Z}), we obtain
\beq
Z(x,y,\D t=1) = \sum_N (2e^{-\g'})^N e^{N L_{(-1)}(u,v,w)} ,
\eeq
where we have defined $e^{-\g'}=b e^{-\g}$ to allow for a multiplicative renormalization
of the partition function at fixed $N$, in analogy with the two-dimensional case
(\ref{2d-sumT}).
The continuum limit is obtained for $u\to u_c$, $v \to v_c$, $w \to w_c$, if we
tune $\g' = \log 2 + L_{(-1)}(u_c,v_c,w_c)$, yielding
\beq
Z(x,y,\D t=1) = \sum_N e^{N [L_{(-1)}(u,v,w)-L_{(-1)}(u_c,v_c,w_c)]} \sim
\frac{1}{L_{-1}(u_c,v_c,w_c)-L_{-1}(u,v,w)} ,
\eeq
for the leading term.
To derive the quantum Hamiltonian in the ``$X$-representation",
we substitute the known function $L_{-1}(u,v,w)$ as well as an ansatz for
the scaling relations into the evolution equation
\beq
\psi(x,t+1)=\oint\frac{dy}{2\p i y} Z(x,y^{-1},\Delta t=1)\psi(y,t)
\eeq
for the wave function
(remember that $x$ and $y$ were absorbed in $u$ and $v$),
and evaluate its continuum limit to order $a$, namely,
\beq \label{H-formula}
 \left(1- a\hat H_X+{\cal O}(a^2)\right)\psi(X,T)=-a^2\int_{-i\infty+\m}^{+i\infty+\m}\frac{dY}{2\p i}
 Z(X,-Y;a)\psi(Y,T).
\eeq
In (\ref{H-formula}), $\m$ is chosen such that the integration contour lies to the right of the
singularities of $\psi(Y,T)$ and
to the left of those of $Z(X,-Y;a)$, around which we have to close it.

To show that to lowest order the identity operator is reproduced and to extract
the Hamiltonian at the next order,
we clearly need the first two orders in $a$ of $L_{-1}(u,v,w)-L_{-1}(u_c,v_c,w_c)$.
As discussed in Sec.\ \ref{replica-sec} above, to get the order-$a^2$ terms
right we need to use the replica trick
up to the second moment approximation, $i.e.$ $L_{-1}=L_2-3 L_1$.
It does not matter that we do not have a closed analytic expression available, since we will
only need the expansion
up to order $a^2$. This can be found perturbatively around the solution at the critical point
(where we {\it can} solve for the eigenvalues of $(A^{\otimes 2}+B^{\otimes 2})/2$), with the
eigenvalue problem perturbed in accordance
with the chosen scaling of coupling constants.

In anticipation of difficulties with the scaling of $k$ we first discuss the results for $k=0$.
Given the analytic expression of $L_1$ and after computing
\beq
L_2(X,Y,G,\L)=\ln \left( \frac{4}{9}+a\frac{4}{9}\sqrt{X+Y}+ \frac{a^2}{27}\left(
\frac{4 X^2 + 5 X Y + 4 Y^2}{X+Y}+\frac{(4b_1+2b_2)\L}{\sqrt{X+Y}} \right)  \right)+O(a^3),
\eeq
we have all we need to calculate the partition function, resulting in
\beq \label{Z-k=0}
a^2 Z(X,-Y;\L)=\frac{2 a}{\sqrt{X-Y}}+a^2\left( \frac{5}{6}+\frac{XY}{(X-Y)^2}-\frac{\L}{(X-Y)^{3/2}}
 \right)+O(a^3),
\eeq
where we have absorbed a finite numerical factor in $\Lambda$.
This expression is in perfect agreement with our previous discussion, since due to the entropy factor
$\NN(A)\sim A^{-1/2} e^{\l_0 A}$ from the boundaries we expect the leading term to have an inverse
square-root singularity if the transfer matrix is to reduce to the identity at lowest order.
Using the fractional derivative method before the integration in (\ref{H-formula}) or, alternatively,
the method of the auxiliary Hamiltonian one finds
\beq \label{H}
\hat H_{\cal A}=-{\cal A}^{\frac{3}{2}}\frac{\partial^2}{\partial {\cal A}^2}
-\frac{3}{2}{\cal A}^{\frac{1}{2}}\frac{\partial}{\partial {\cal A}}
-\frac{1}{16}\frac{1}{{\cal A}^{\frac{1}{2}}}+\L \ {\cal A}
\eeq
for the quantum Hamiltonian in the ``${\cal A}$-representation" (with $\cal A$ denoting
the (finite) continuum area). This looks like a bona-fide Hamiltonian, with a second-order
kinetic term and a potential term depending on the cosmological constant.
In the absence of a dimensionful
Newton coupling $G$, the factor ${\cal A}^{\frac{1}{2}}$ multiplying the kinetic term must be
present for dimensional reasons.
Note also the appearance of the factor $3/2$ in front of the first-order derivative, ensuring
the self-adjointness of the Hamiltonian with respect to the trivial $d{\cal A}$-measure.
By introducing a new variable
\beq \label{L-variable}
L=4\sqrt{{\cal A}}
\eeq
and simultaneously defining new wave functions\footnote{With such a change of wave functions
we make sure that the measure is preserved, in the sense that
\beq
\int d{\cal A}\ \psi_1({\cal A})\psi_2({\cal A}) = \int dL\ \phi_1(L) \phi_2(L) \nonumber.
\eeq
Note that we also require
\beq
\int d{\cal A}\ \psi_1({\cal A})\hat H_{\cal A} \psi_2({\cal A}) = \int dL\ \phi_1(L)\hat H_L \phi_2(L) ,\nonumber
\eeq
which implies
\beq
\hat H_L=\hat H_{{\cal A}\rightarrow L}+\sqrt{L}\left[\hat H_{{\cal A}\rightarrow L},\frac{1}{\sqrt{L}}\right],
 \nonumber
\eeq
with the $\cal A$ in the Hamiltonian substituted according to (\ref{L-variable}).          }
\beq \label{L-function}
\phi(L)=\frac{\sqrt{2 L}}{4} \psi(\frac{L^2}{16}),
\eeq
the Hamiltonian becomes
\beq \label{H-L}
\hat H_L=-L \frac{\partial^2}{\partial L^2}-\frac{\partial}{\partial L}+\frac{\L}{16} L^2 ,
\eeq
with the $1/L$-term having disappeared from the potential.
Apart from the cosmological term, which has the appropriate dimension for a
(2+1)-dimensional Hamiltonian,
this is exactly the quantum Hamiltonian of the (1+1)-dimensional CDT
model \cite{2d-ldt} (see also \cite{difra-calogero}
for a similar transformation of variables).

Since by setting $k=0$ we have not included any spacetime-curvature
term in the discrete action, it is clear that the kinetic terms in (\ref{H}) and (\ref{H-L})
have their origin in the ``entropy" of configurations, or, using a continuum language,
in the non-trivial path integral measure underlying the dynamically triangulated model.
This is also underlined by the absence of such a kinetic term from a related (1+2)-dimensional
model considered and solved in \cite{difra2}, which uses the same product structure as our model,
but works instead with a {\it fixed}, flat base manifold.
Rather intriguingly, this gravity-inspired model (for finite $t$) can be related through
an inversion formula to a problem of hard hexagons on a regular triangulation.
Its one-step propagator can be seen to resurface in our model as contributing just
one of the terms in the partition function, namely, $\tilde Z_N=1/Tr(AB)^{N/2}$.
Solving the model in the large-$N$ limit does not require the replica trick, and simply leads to
$\tilde Z_N\sim 1/\l_{max}^{N/2}$, where $\l_{max}$ is the largest eigenvalue of the matrix $AB$.
It is straightforward to extract the Hamiltonian, which only contains
a term proportional to the area, and no derivatives, which implies that the area is not a dynamical
quantity in the continuum
limit\footnote{The mapping to the hard hexagon model of \cite{difra2} suggested
a fractal dimension $d_f=12/5$ for the (1+2)-dimensional simplicial complexes.
The relation with our results for the associated one-step propagator
is currently unclear. We conjecture that in our model, where we sum over all base triangulations,
the fluctuations of the base serve as a stabilizer for the geometry and lead to an effective,
non-anomalous dimension $d_f=3$ for finite times.
}.

Returning to the analysis of our model,
the case with $k\neq 0$ raises issues similar to those already encountered when
computing the spacetime volume and curvature.
The $a/G$-term dominates the continuum limit, giving rise to
\beq
a^2 Z(X,-Y;G,\L)=\frac{2 a^{3/2}}{3\sqrt{(c_2-2c_1)/G}}+
 a^2 \frac{(6c_1-5c_2+8c_3)/G-4\sqrt{4X-4Y-3(c_2/G)^2}}{(36c_1-18c_2)/G}+O(a^{5/2}),
\eeq
which is difficult to interpret within the scheme of the previous section (and to this order does not
even contain any reference to the cosmological constant $\Lambda$!)
Things simplify somewhat once we choose $c_2=2c_1$,
in which case we obtain
\beq
\label{zxyg}
a^2 Z(X,-Y;G,\L)=\frac{2 a}{4 c_3-c_2+\sqrt{X-Y-\frac{3}{4}\left(\frac{c_2}{G}\right)^2}}+
 O(a^2) .
\eeq
We have not bothered to include the next term in the expansion, because already
the lowest-order expression gives nothing like the
desired form ${\cal N}({\cal A})\delta({\cal A}-{\cal A'})$ upon performing an inverse
Laplace transform. It is not totally inconceivable that the offensive terms in
(\ref{zxyg}) could still be ``argued away". Note that we have encountered the numerical
term $(4 c_3-c_2)$ previously in the expressions (\ref{rav1}) and (\ref{rav2}) for the average
curvature. In particular, {\it if} we had a good reason for why $(4 c_3-c_2)$ could be set
to zero (which we currently do not), it would imply that in the case $k=0$ the average curvature was
turned into a finite expression. However, even if  $(4 c_3-c_2)$ could be made to vanish, it would still
leave us with the problem of having to absorb the shift in the location of the singularity
from $Y=X$ to $Y=X-\frac{3}{4}\left(\frac{c_2}{G}\right)^2$. In terms of the ${\cal A}$-representation,
the shift can be related to an overall multiplicative operator, which replaces the simple
transfer matrix $e^{-a\hat H_{\cal A}}$
by $e^{-a\hat H_{\cal A}}e^{\frac{3}{4}\left(\frac{c_2}{G}\right)^2\hat{\cal A}}$.
This might be related to a redundancy among the renormalized couplings of
the model inherent in the ansatz (\ref{canonical}), (\ref{canonical2}), but we have not
yet been able to make the relation precise.

\section{Conclusions} \label{sec-conclusions}

In this paper, we have for the first time derived a continuum Hamiltonian from a
three-dimensional quantum gravity model in terms of Causal Dynamical Triangulations.
This was made possible by restricting ourselves to a subclass of all triangulations,
which possess a product structure in the spatial direction, in addition
to the usual product structure in the time direction underlying the causal nature of
the model. This restriction enabled us to apply a number of analytical methods,
including the inversion formula, a random matrix formulation and the replica trick, to
solve the one-step propagator with free boundary conditions and then perform its
continuum limit. This led to an explicit expression for the quantum Hamiltonian
(\ref{H}) in ${\cal A}$-space, with wave functions depending on the area ${\cal A}$
of two-dimensional spatial slices.

Taking into account the special, topological character of three-dimensional pure
gravity, we have argued that for obtaining the full dynamics of the theory,
it is sufficient to keep track of only a finite number of
boundary data when deriving the model's transfer matrix from the
one-step propagator, and sum over everything else. This should not be
confused with an approach where the remaining degrees of freedom
(essentially the conformal mode of the metric) are fixed at the outset.
In our model, all of these are still present, but are summed over in the
discrete path integral. For simplicity, what we have done in the present piece of work is
to keep track of only a single variable, the volume (area) of the two-dimensional
universe. Because of the cylindrical topology of our spatial slices, we
expect there to be one additional Teichm\"uller parameter, corresponding to
the ratio between the area of the cylinder and the length of the cycle (the
boundary of the cylinder). To include this parameter in our model will
require to separately keep track of the number of matrices $A$ and $B$
(for the incoming and outgoing cylinder) in the traces, instead of only their
sum $N$, as we have done here. This task is in principle solvable
with the help of the so-called microcanonical method \cite{paladin},
based on an idea similar to that of the replica trick. This issue is currently
under consideration, with some progress reported in \cite{thesis}.

The fact that we have derived a non-trivial three-dimensional dynamics does
{\it a posteriori} justify our restriction to a subclass of all three-dimensional Lorentzian
triangulations. Since the continuum limit we have identified was obtained
by setting the bare inverse Newton constant to zero, it is clear that the
resulting continuum Hamiltonian describes the ``collective" effect of the
quantum fluctuations we have summed over in the path integral, and
which elsewhere we have called the entropy of the model. In particular,
the net effect of these fluctuations is to generate a kinetic term for
the area which makes the Hamiltonian {\it bounded below}, and therefore
has the opposite sign from the corresponding term for the ``global conformal
mode" in the gravitational action. This further supports a mechanism
already observed elsewhere in CDT models in three and four dimensions,
namely, the presence of contributions from the path integral measure
compensating for the divergence due to the conformal mode of the Euclidean
gravitational action.

There is a (small) price we are paying for using the additional constraint
of a double product structure for the triangulations, which after all was not
motivated by physics, but simply by the desire to be able to apply a number
of special solution techniques. The application of the inversion formula
implies that we are evaluating first the limit of infinitely many building blocks in
{\it one} of the spatial directions (the ``height" of the towers), and only then
the corresponding limit in the complementary direction, corresponding to
what we have called the large-$N$ limit. The effect of taking these two
limits in sequence, and not simultaneously, is that the $u$-$w$-diagram
(Fig.\ \ref{zeros}), which describes the behaviour {\it after} taking the first
of the infinite sums, has only a single point at which an infinite-volume
limit can be defined. This is in contrast to the corresponding phase diagram
for three-dimensional CDT, which has an entire critical line $\lambda_c (k)$
along which the infinite-volume limit can be taken.

The analytic solution found here has also highlighted a problem with the
scaling of Newton's constant, which has already been encountered in
previous analyses of the phase structure of three-dimensional CDT
models \cite{ABAB3}. The issue is that a standard canonical, additive
renormalization of Newton's constant as in (\ref{canonical}) does not seem
to lead to a good
continuum limit, simply because the term containing the renormalized Newton
constant $G$ has the lowest order in $a$, and dominates everything else.
The only way so far in which we have managed to obtain a nontrivial continuum limit
was by setting $k=0$, which implies that the
Newton constant does not play a dynamical role, which is of course
compatible with the absence of local gravitational excitations in three
dimensions. We believe a deeper understanding of the relation between
certain scalings of the coupling constants and that of physically relevant
quantities, like the curvature discussed in Sec.\ \ref{sec-continuum} above
would be useful, and potentially relevant for a better understanding of
the analogous issues in four dimensions. --
Even if this and other issues still remain open, we think that the work
presented here presents an important
step in the analytical understanding of CDT models in $d>2$ dimensions.

\vspace{.7cm}

\noindent {\bf Acknowledgements.} We thank R.\ Costa-Santos for discussions during
the early stages of the project. D.B. and R.L. are partially supported
through the European Network on Random Geometry
ENRAGE, contract MRTN-CT-2004-005616. R.L. acknowledges
support by the Netherlands Organisation for Scientific Research
(NWO) under their VICI program.
D.B. thanks the Laboratoire de Physique Th\'eorique (LPT) of the ENS in Paris, and the
Service de Physique Th\'eorique (SPhT), CEA Saclay,
and F.Z. thanks the Institute for Theoretical Physics of Utrecht University for hospitality
during some stages of the preparation of this work.

\clearpage

\renewcommand{\thesection}{A-\Roman{section}}
% redefine the command that creates the section no.
\setcounter{section}{0}  % reset counter
\renewcommand{\theequation}{A-\arabic{equation}}
% redefine the command that creates the equation no.
\setcounter{equation}{0}  % reset counter
\section{The discrete action} \label{App-action}

We derive in this appendix the exact expression for the action of our model.
All the definitions and necessary ingredients were already introduced in \cite{3d-ldt}, here we just
recall and use them to derive the precise expression for our particular case.
We start from the Einstein-Hilbert action plus a Gibbons-Hawking boundary, that is,
\beq
{\cal S}=\int_M d^3x\sqrt{g(x)}\left(\L-\frac{R(x)}{2G}\right)+\frac{1}{G}\int_{\partial M}
d^2x\sqrt{h(x)} K(x),
\eeq
and
then use Regge's prescription for the corresponding quantities on a simplicial
manifold \cite{regge,hartle} to obtain the discrete action
\beq \label{S-Regge}
{\cal S}\rightarrow S=\L_b\sum_{\s_3} V_{\s_3}-\frac{1}{G_b}\sum_{\s_1\in \dot M} V_{\s_1}
   (2\pi-\sum_{\s_3\supset\s_1}\th_{\s_3\rhd\s_1})-\frac{1}{G_b}\sum_{\s_1\in \partial M} V_{\s_1}
   (\pi-\sum_{\s_3\supset\s_1}\th_{\s_3\rhd\s_1}),
\eeq
where $\s_n$ is an $n$-dimensional simplex, $V_{\s_n}$ its volume, $\dot M$ is the interior of the
simplicial manifold and $\partial M$ its boundary, $\th_{\s_3\rhd\s_1}$ is the dihedral angle
of $\s_3$ at $\s_1$, and the subscript $b$ on the coupling constants stands for {\it bare}.

In Dynamical Triangulations things simplify considerably because the edge lengths
are held fixed. In the old, Euclidean models
one would have only a single type of building block in any dimension,
since all the edge lengths were chosen equal.
In CDT the Lorentzian nature of the model allows us to have two different
types of edge lengths, corresponding to time- and space-like directions.
Because of the way the causal gluing rules are implemented, what used to be
a space-like and what used to be a time-like link can still be distinguished after
having mapped a Lorentzian triangulation to its Euclidean counterpart.
Let us define the ratio $r=l_t^2/l_s^2$, where $l_t^2$ denotes the squared edge length
for all time-like links, and $l_s^2$ that for all space-like ones.
Its allowed range in the Euclidean signature is $r>\ee>0$, where $\ee$ is a positive
constant, which is the lower bound of
the triangular inequalities and depends on the number of dimensions
$d$ ($\ee=1/4$ for $d=2$,
$\ee=1/2$ for $d=3$ and $\ee=7/12$ for $d=4$).
An analytic continuation of $r$ in the complex plane
to negative values defines the (inverse) ``Wick rotation" in CDT.
Throughout this work, we will stick to positive values for $\epsilon$, corresponding
to a Euclideanized, real partition function.

In (2+1) dimensions we have precisely two different kinds of building blocks,
tetrahedra with three space-like and three time-like edges, and tetrahedra with
two space-like and four time-like edges.
We will use the notation $N_{ij}$ for the number of $(i,j)$-simplices of dimension $i+j-1$
having $i$ vertices in a constant-time slice $t$ and $j$ vertices in the subsequent one
at time $t+1$. The Regge action (\ref{S-Regge}) in this case becomes
\beq
\begin{split}
S= &\L_b a^3 \left( V_{2,2} N_{2,2} + V_{3,1} (N_{3,1}+N_{1,3})\right)
-\frac{a}{G_b}\bigg(2\pi\sqrt{r} N_{1,1}^{(\dot M)}+\pi \sqrt{r} N_{1,1}^{(\partial M)}+\\
 &-\sqrt{r}\left[4 \th_{2,2}^t N_{2,2} +3 \th_{1,3}^t (N_{1,3}+N_{3,1}) \right]
  +\pi (N_{2,0}+N_{0,2})+\\
 &-\left[2 \th_{2,2}^s N_{2,2} +3 \th_{1,3}^s (N_{1,3}+N_{3,1}) \right]\bigg),
\end{split}
\eeq
where we have distinguished the dihedral angles at time-like and space-like links
by the superscripts $t$ and $s$. Relevant volumes and angles are easily found to be
\beq
V_{2,2} = \frac{\sqrt{2r-1}}{6\sqrt{2}},\  \ \ \  V_{3,1} = \frac{\sqrt{3r-1}}{12} ,
\eeq
\beq
\begin{split}
&\cos \th_{1,3}^s = \frac{1}{\sqrt{3}\sqrt{4r-1}}, \ \ \ \cos \th_{1,3}^t = \frac{2r-1}{4r-1}, \\
&\cos \th_{2,2}^s =  \frac{4r-3}{4r-1}, \ \ \ \cos \th_{2,2}^t = \frac{1}{4r-1}.
\end{split}
\eeq
Finally, using Euler's formula and the Dehn-Sommerville relations for triangulated
manifolds with
boundary\footnote{Note that the boundary of our ``sandwiches" of three-dimensional
spacetime (the product of a finite cylinder with an interval), which has space-like as well
as time-like components, is connected and has the topology of a torus.},
we arrive at the expression
\beq\label{}
  S=\alpha (N_{13}+N_{31}) + \beta N_{22} +\gamma N
\eeq
for the discrete action, where $\alpha$, $\beta$ and $\gamma$ depend on the
dimensionless bare
coupling constants $\l$ and $k$ according to
\beq\label{alphabeta2}
\begin{split}
  &\alpha = \left(-\pi\sqrt{r}+3 \sqrt{r}\arccos \frac{2r-1}{4r-1}
            -\frac{3}{2}\pi+3\arccos \frac{1}{\sqrt{3}\sqrt{4r-1}}\right)k
   +\frac{\sqrt{3r-1}}{12}\lambda =-c_1 k + b_1 \lambda, \\
  &\beta = \left(-2\pi\sqrt{r} +4 \sqrt{r}\arccos \frac{1}{4r-1}+2\arccos \frac{4r-3}{4r-1}\right)k
   +\frac{\sqrt{2r-1}}{6\sqrt{2}}\lambda = c_2 k + b_2 \lambda, \\
  &\gamma = \left(6 \pi \sqrt{r}-\pi\right) k = c_3 k .
\end{split}
\eeq

\section{The inversion formula} \label{App-inversion}

\begin{figure}[h]
\centering
\includegraphics[width=12cm]{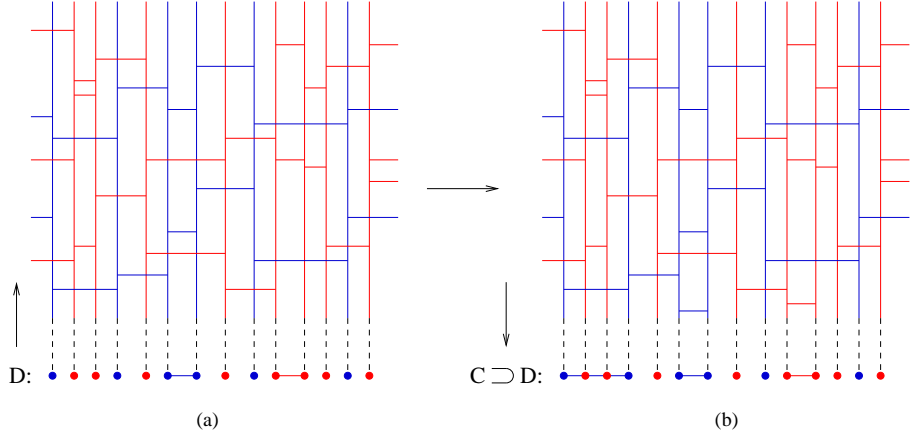}
\vspace*{13pt}
\vspace*{13pt}
\caption{\footnotesize By completing an arbitrary triangulation of our sandwich
geometry with a hard-dimer
configuration $D$ (a),
we build a larger triangulation (b) whose projection $C$ contains
$D$. With this procedure (with fixed $D$) we build
all triangulations whose projection $C$ contains $D$ exactly once.}
\label{color-completion}
\end{figure}

The proof of the inversion formula in (2+1) dimension (for a fixed sequence
$S_N$ of $N$ blue and red towers) proceeds very
similarly to that in (1+1) dimensions given in \cite{difra2}.
We first switch to the dual picture of a triangulation, and assign a weight
$u$ or $v$ to each horizontal edge, depending on its colour, and
a weight $w$ to each red-blue intersection, as depicted in Fig. \ref{3d-dual} above.
\begin{figure}[h]
\centering
\includegraphics[width=6cm]{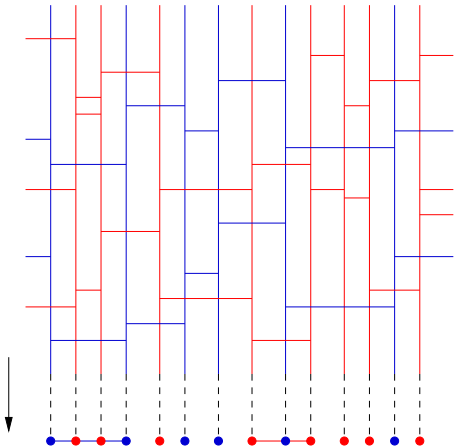}
\vspace*{13pt}
\caption{\footnotesize Decomposing the dual picture of a sandwich geometry
into a sequence of projections onto its one-dimensional base line.
The projection is obtained by letting the
``bottom layer" of the heap of pieces, consisting of its lowest-lying
edges, drop down to the horizontal line at the base.
Each such projection defines a hard-dimer configuration on the horizontal
line.}
\label{color-projection}
\end{figure}
Next, we decompose a given configuration into a sequence of
projections onto the base, as illustrated in Fig.\ref{color-projection}.
Such a projection gives rise to a hard-dimer configuration, if by a dimer
we simply mean an edge linking two nearest vertices of the same colour.
The ``hardness" refers to the fact that any vertex of the base can be occupied by
at most one dimer, incorporating a mutual avoidance of dimers.
Such a construction is possible because the dual graph is a heap of pieces.
One may think of the vertical lines in Fig.\ \ref{color-projection} as tracks
along which the horizontal edges can slide up and down,
with the only restriction of not being allowed to touch or pass each other.
Using this decomposition, we can write the partition function $Z_T(t)$ as
\beq \label{Zc}
 Z_{S_N}(u,v,w)=\sum_{{\rm hard}\ {\rm dimer}\ {\rm config.}\ C}
u^{|C|_b} v^{|C|_r} w^{|\cap C|}  Z_{S_N}^{(C)}(u,v,w),
\eeq
where the sum extends over all hard-dimer configurations $C$ on the
one-dimensional lattice (including the empty configuration).
The numbers $|C|_b$ and $|C|_r$ count the blue and red dimers in $C$,
and $|\cap C|$ the crossings between dimers and sites of different colour
in the configuration $C$. For fixed $C$, $Z^{(C)}(u,v,w)$
is the restricted partition function involving those configurations having
projection $C$, and from which we have factored out the weight
$u^{|C|_b} v^{|C|_r} w^{|\cap C|}$
of the projected part. More generally, we have the relations
\beq \label{Zd}
   u^{|D|_b} v^{|D|_r} w^{|\cap D|} Z_{S_N}(u,v,w)=\sum_{C \supset D}
u^{|C|_b} v^{|C|_r} w^{|\cap C|} Z_{S_N}^{(C)}(u,v,w),
\eeq
valid for any hard-dimer configuration $D$ (eq.\ (\ref{Zc}) corresponding to $D=\emptyset$).
This expresses the fact that by completing any dual geometric configuration
under consideration by a given row of horizontal edges (corresponding to a
hard-dimer configuration $D$), one builds each configuration
having a projection containing $D$, i.e. having $D$ as a sub-configuration
exactly once, see Fig.\ref{color-completion}.
Let us now rewrite (\ref{Zd}) as
\beq \label{g=f.zeta}
  g(D)=\sum_C f(C)\zeta(C,D)
\eeq
with
\beq
  g(D)=u^{|D|_b} v^{|D|_r} w^{|\cap D|} Z_{S_N}(u,v,w),
\eeq
\beq
  f(C)=u^{|D|_b} v^{|D|_r} w^{|\cap D|} Z_{S_N}^{(C)}(u,v,w),
\eeq
and
\beq
  \zeta(C,D)=
  \begin{cases}
  1 & \text{if $C\supset D$},\\
  0 & \text{otherwise}.
  \end{cases}
\eeq
Thinking of the space of configurations as a vector space in this manner,
(\ref{Zd}) can be thought of as a vector-matrix multiplication,
with an upper-triangular matrix $\zeta(C,D)$ and with all diagonal elements
being equal to 1. Then $\zeta$ has an inverse matrix $\mu(D,C)$ of the form
\beq
  \mu(D,C)=
  \begin{cases}
  (-1)^{|D|-|C|} & \text{if $D\supset C$},\\
  0 & \text{otherwise}.
  \end{cases}
\eeq
This property can be verified by noting that
\beq
  \sum_D \zeta(C,D)\mu(D,C')=
  \begin{cases}
  \sum_{D\supset C'\atop D\subset C}(-1)^{|D|-|C|} & \text{if $C\supset D\supset C'$},\\
  0 & \text{otherwise},
  \end{cases}
\eeq
and
\beq
  \sum_{D\supset C'\atop D\subset C}(-1)^{|D|-|C|}=
  \begin{cases}
  1 & \text{if $C=C'$},\\
  \sum_{i=0}^{|C|-|C'|}(-1)^i\matrice |C|-|C'|\\i\ematrice = (1-1)^{|C|-|C'|}=0 & \text{otherwise}.
  \end{cases}
\eeq
We can now invert (\ref{Zd}) as\footnote{This is the famous M\"{o}bius inversion
formula of the theory of partially ordered sets, and $\mu$ is the associated M\"{o}bius function.}
\beq
  u^{|C|_b} v^{|C|_r} w^{|\cap C|} Z_{S_N}^{(C)}(u,v,w) = \sum_{D \supset C} (-1)^{|D|-|C|}
  u^{|D|_b} v^{|D|_r} w^{|\cap D|} Z_{S_N}(u,v,w).
\eeq
Noting that $Z_{S_N}(u,v,w)$ factors out of the sum on the right-hand side,
we finally get
\beq
  Z_{S_N}(u,v,w)=
   \frac{(-u)^{|C|_b} (-v)^{|C|_r} w^{|\cap C|} Z_{S_N}^{(C)}(u,v,w)}
{\sum\limits_{D \supset C} (-u)^{|D|_b} (-v)^{|D|_r} w^{|\cap D|}},
\eeq
where we have used that  $|D|=|D|_b+|D|_r$, leading to the minus sign
in front of both $u$ and $v$.
Picking $C=\emptyset$, we arrive at the fundamental inversion relation (eq.\ (\ref{fund3}) above)
\beq
  Z_{S_N}(u,v,w)= \frac{ 1}{Z_{S_N}^{hcd}(-u,-v,w)},
\eeq
where
\beq
  Z_{S_N}^{hcd}(u,v,w)= \sum_{{\rm hard}\ {\rm dimer}\ {\rm config.}\ D}
  u^{|D|_b} v^{|D|_r} w^{|\cap D|}
\eeq
denotes the partition function for hard coloured dimers with fugacity $u$ ($v$) per blue (red) dimer
and weight $w$ per crossing.

\section{Numerical computations} \label{App-numeric}

\begin{figure}[t]
\centering
\includegraphics[width=10cm]{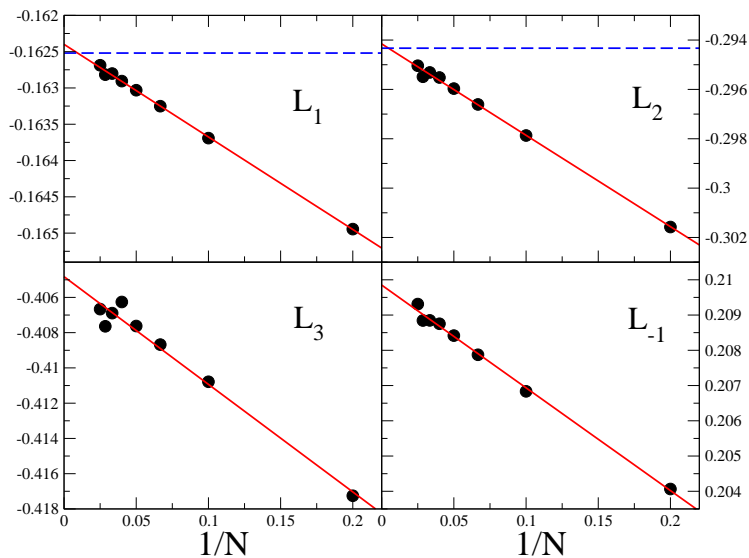}
\caption{
The numerical results for $L_{n}(N)$, $n=-1,1,2,3$, as a function of $1/N$ for $u=0.24$ and $w=0.1$.
The full lines are fits to $L_{n}(N) = L_{n} + A_n/N$. The dashed lines are the analytical values for
$L_1$ and $L_2$ computed using the replica trick. They differ from the extrapolation by about $0.1\%$.
}
\label{1suN}
\end{figure}

A direct numerical computation of the generalized Lyapunov exponents $L_n$, defined
in (\ref{lns}) above, is possible following the procedure
given in \cite{crisanti}. One extracts a large number of values of $\g$ at fixed $N$,
computes the average $\langle e^{n \g N} \rangle$, and plots
$L_n(N)=N^{-1} \ln \langle e^{n \g N} \rangle$ as a function of $1/N$. Usually
$L_n(N)$ turns out to be linear in $1/N$ for large $N$,
\beq
L_n(N) \sim L_n + A_n/N,
\eeq
which allows for a good extrapolation of the data to $N= \io$, see Fig.~\ref{1suN}
for an example.
The values of $\g$ are extracted as follows \cite{crisanti}:
\begin{enumerate}
\item construct a random unimodular
vector $v$ (e.g. by extracting the components from a Gaussian distribution);
\item extract
the numbers $q_j$ and apply the matrix $M_{q_j}$ to $v$ to obtain a vector $v' = \prod_j M_{q_j} v$;
\item compute $|v'| \simeq e^{N \g}$, which holds to leading order since the modulus of $v'$ is dominated
by the same largest eigenvalue of $\prod_j M_{q_j}$ which also dominates $\Tr \prod_j M_{q_j}$.
\end{enumerate}
By repeating this procedure with random choices of the $\{q_j\}$,
we obtain a set of $\NN$ values $\g_i$, from which
we can construct a histogram $\pi(\g)$
and extract the function $S(\g)$ at given $N$, see Fig.~\ref{Sgamma}. The coincidence of the
curves for different $N$ indicates that the asymptotic limit $N \to \io$ has been reached.
To compute $L_n$, we first have to compute the average of
$\g$, $\bar\g = \frac1\NN \sum_{i=1}^\NN \g_i$,
and then use
\beq
e^{N L_n(N)} =
\langle e^{n \g N} \rangle = e^{n \bar\g N} \frac1\NN \sum_{i=1}^\NN e^{n N (\g_i-\bar \g)} \hskip10pt
\Leftrightarrow \hskip10pt
L_n(N) = n \bar\g + \frac1N \ln \left[\frac1\NN \sum_{i=1}^\NN e^{n N (\g_i-\bar \g)} \right] .
\eeq
By subtracting $\bar \g$ we can compute the average as long as the quantity
$N (\g_i - \bar \g) \sim \sqrt{N}$ is not too large. In practice we are limited to $N < 50$,
but this is enough to get a very good linear extrapolation to the limit $N \to \io$.
The difference between the numerical computation and the exact analytic expressions
for $L_1$ and $L_2$ is of the order of $10^{-4}$, see Fig.~\ref{1suN}.
We believe that a similar error affects the computation of $L_{-1}$ too.

\section{Fractional derivatives and inverse Laplace transforms} \label{App-fracder}

Although the notion of {\it fractional calculus} may at first appear
somewhat extravagant,
this is an old and well-studied subject in mathematics, with plenty of
applications,
as testified by the number of books on the subject (see, for instance, the
list of references
in \cite{mainardi,podlubny}). At the heart of the theory of fractional
calculus is the
definition and study of
two operators $J^{\a}:f(x)\rightarrow J_x^{\a}f(x)$ and
$D^{\a}:f(x)\rightarrow D_x^{\a}f(x)$
on a sufficiently large class of functions $\{ f(x)\}$ and for a positive
real number $\a$,
with the following properties.
\begin{enumerate}
\item When $\a=n$ is a positive integer, the operator $J_x^{\a}$ gives the
same result as $n$-fold
integration, and $D_x^{\a}$ gives the same result as the usual $n$-th
derivative
$\frac{d^n}{dx^n}$.
\item The operators of order $\a=0$ are the identity operator.
\item $J_x^{\a}$ and $D_x^{\a}$ are linear operators.
\item For any $\a$, $\b$, the semigroup property holds for $J_x^{\a}$,
namely,
$J_x^{\a}J_x^{\b}f(x)=J_x^{\a+\b}f(x)$.
\end{enumerate}

There are many inequivalent definitions of fractional derivative/integral
satisfying
these properties, which we are not going to review here. Instead, we will
recall the definition and properties of the one that turns out to be
relevant for our purposes, the so-called Weyl fractional
derivative/integral.
(Reference \cite{osler} has a short review, which also contains a clear
explanation
for why the various  definitions are inequivalent.)
The Weyl fractional integral is defined as\footnote{The subscript
``$\infty$" refers to the
extremum of integration.
Other choices of extremum would give rise to different (and inequivalent)
definitions of
fractional derivative/integral, which we would denote by $J_{x,x_0}^{\a}$,
$D_{x,x_0}^{\a}$.}
\beq \label{weyl-int}
J_{x,\infty}^{\a}f(x)=\frac{1}{\G(\a)}\int_x^{\infty}dt f(t)(t-x)^{\a-1}.
\eeq
The natural definition of the derivative operator would be to take
$D_{x,\infty}^{\a}=J_{x,\infty}^{-\a}$
but this is divergent. A standard trick is then to define
\beq \label{weyl-der}
D_{x,\infty}^{\a}f(x)=\frac{d^{[\a ]+1}}{dx^{[\a ]+1}}\ J_{x,\infty}^{[\a
]+1-\a}f(x),
\eeq
where $[\a ]$ is the integer part of $\a$. The rationale behind this
definition is that
in this way $D_x^{\a}$ is the left inverse of $J_x^{\a}$, $i.e.$
$D_x^{\a}J_x^{\a}f(x)=f(x)$, the
same as when $\a =n$ is an integer.
With this definition it can be checked that
\beq \label{frac-exp}
D_{x,\infty}^{\a}e^{-Ax}=(-1)^{[\a ]+1}A^{\a}e^{-Ax},
\eeq
which, apart from the sign, is exactly what we need, as explained in the
main text.
According to the analysis there, we need to compute
\beq
\tilde Z(x,y,\Delta t=1)=D_{\ln x,\infty}^{1/4} D_{\ln y,\infty}^{1/4} Z(x,y,\Delta t=1)\ ,
\eeq
which in the continuum limit we can write as (see Sec.~\ref{sec-hamiltonian})
\beq
\tilde Z(X,Y)=D_{a^2 X,\infty}^{1/4} D_{a^2 Y,\infty}^{1/4} a^2 Z(X,Y)=
a\ D_{X,\infty}^{1/4} D_{Y,\infty}^{1/4} Z(X,Y) .
\eeq
To leading order in $a$, we expect the transfer matrix to be given by the
delta function $\d (A_1-A_2)$, which implies that we
can simply take a $\frac{1}{2}$-derivative with respect to one of the arguments.
From expression (\ref{Z-k=0}) for $Z(X,Y)$ we have
\beq
\begin{split}
a D_{X,\infty}^{1/2} Z_{l.o.}(X,Y) &=D_{X,\infty}^{1/2} \frac{2}{\sqrt{X+Y}}=
 \lim_{X_0\to\infty}\frac{d}{dX}\frac{1}{\G(1/2)}\int_X^{X_0}dt \frac{2}{\sqrt{t+Y}}\frac{1}{\sqrt{t-X}}\\
 &=-\frac{2}{\sqrt{\pi}}\frac{1}{X+Y} ,
\end{split}
\eeq
at leading order ($l.o.$), which is proportional to the Laplace transform of the delta function.
Note that the limit has to be taken after the derivative, since otherwise it would be divergent.
This is the kind of convergence problem mentioned in Sec.~\ref{sec-gluing} (which by the
way would remain if we used two $\frac{1}{4}$- instead of one
$\frac{1}{2}$-derivative).
It is not a real problem because all we need is an operator which gives the same result
as (\ref{frac-exp}) when acting on exponential functions. Since for exponential functions
the limit can be taken both before and after the derivative, we have a freedom in choosing
the order of these operations.

For the cosmological term in (\ref{Z-k=0}), interchanging limit and derivative is
unproblematic and leads to
\beq
 D_{X,\infty}^{1/4} D_{Y,\infty}^{1/4} \frac{-a\L}{(X+Y)^{3/2}}=-\frac{2 a}{\sqrt{\pi}}\frac{1}{(X+Y)^2} ,
\eeq
which is proportional\footnote{Note that although we obtained
the same proportionality factor $-2/\sqrt{\pi}$ for both the identity and the cosmological term,
there is a relative sign between the two because in the first case we used only one derivative,
resulting in a minus sign (from the minus sign in (\ref{frac-exp})). If this sign difference
was not there, we would have a negative effective cosmological constant.}
to the Laplace transform of $a\L A_1 \d (A_1-A_2)$.

\begin{figure}[h]
\centering
\includegraphics[width=6cm]{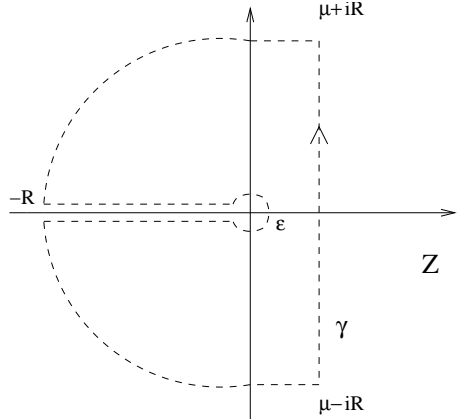}
\vspace*{13pt}
\caption{\footnotesize The Bromwich-Hankel contour used for the inverse Laplace transform.
Integrating the function $e^{Z A_1}/\sqrt{Z}$ along this contour,
we obtain (\ref{cut-integral}) in the limit $R\to\infty$, $\ee\to 0$.}
\label{contour}
\end{figure}

For the remaining term in (\ref{Z-k=0}) we again must take the limit after the derivative,
yielding
\beq
\begin{split}
 D_{X,\infty}^{1/4} D_{Y,\infty}^{1/4} a\left( \frac{5}{6}+\frac{XY}{(X+Y)^2}\right)
 =a \frac{\sqrt{\pi} (X^2-10 X Y+Y^2)}{16 (X+Y)^{5/2}} ,
\end{split}
\eeq
with no contribution from the constant term.
The inverse Laplace transform of this term gives another divergence. This may be avoided by an integration
by parts, but we prefer to take another route and use the second method presented in Sec.~\ref{sec-gluing}.
This consists in performing the inverse Laplace transform on $Z(X,Y)$ directly, without use
of fractional derivatives, and then recognizing the contribution of the entropy in the Hamiltonian.
Let us first see how this works for the leading-order term.
We have to compute the integral
\beq
\begin{split}
I(A_1,A_2) &\equiv \int_{-i\infty+\m}^{+i\infty+\m}\frac{dX}{2\pi i}e^{X A_1}
 \int_{-i\infty+\m}^{+i\infty+\m}\frac{dY}{2\pi i}e^{Y A_2}
 \frac{2}{\sqrt{X+Y}}=\\ &= \int_{-i\infty+\m}^{+i\infty+\m}\frac{dY}{2\pi i}e^{Y (A_2-A_1)}
  \int_{-i\infty+\m}^{+i\infty+\m}\frac{dZ}{2\pi i}e^{Z A_1}
 \frac{2}{\sqrt{Z}}=\\
 &=\d(A_2-A_1) \int_{-i\infty+\m}^{+i\infty+\m}\frac{dZ}{2\pi i}e^{Z A_1}
 \frac{2}{\sqrt{Z}}  .
\end{split}
\eeq
Because of the square root, the integrand has a branch-cut along the semi-axis of
negative $Z$.
The usual Bromwich contour for the Laplace inversion therefore has
to be continued along a Hankel contour
around the cut, as shown in Fig.~\ref{contour}.
We have
\beq \label{cut-integral}
\begin{split}
I(A_1,A_2) &= \d(A_2-A_1) \left( -\int_{-\infty^+}^{0^+}\frac{dZ}{2\pi i}e^{Z A_1}
 \frac{2}{i\sqrt{-Z}}- \int_{0^-}^{-\infty^-}\frac{dZ}{2\pi i}e^{Z A_1}
 \frac{2}{-i\sqrt{-Z}}\right)=\\
 &=\d(A_2-A_1) \int_{0}^{+\infty}\frac{dZ'}{\pi}e^{-Z' A_1}
 \frac{2}{\sqrt{Z'}}=\frac{2}{\sqrt{\pi}}\frac{\d(A_2-A_1)}{\sqrt{A_1}}  ,
\end{split}
\eeq
in which we can recognize the subleading behaviour of (\ref{entropy-A}).
Multiplying by $\sqrt{A_1}$,
we obtain exactly the same result as from the fractional derivative (apart from the sign arising in
(\ref{frac-exp})).

For the cosmological term we have again a convergence problem,
because instead of the square root in
the denominator we have a power of $3/2$, which diverges too fast at zero. This can be removed
by a formal integration by parts in the previous step, and the final result coincides
with that of the fractional derivative.

The remaining terms instead do not cause any problems.
The constant term gives a term proportional to $\d(A_1)\d(A_2)$, which is a non-propagating
and non-universal term of the kind already familiar from two dimensions
(see \cite{ABAB1} and references
therein), and which furthermore gives just zero when multiplied by the inverse entropy factors
$A_1^{1/4}A_2^{1/4}$, in agreement with the fractional derivative result.
The last term has no branch-cut, and is the usual term encountered in two dimensions (see,
for example, \cite{difra-calogero}). After an inverse Laplace transform it leads to an
expression
\beq
\hat H''_{kin}\d(A_2-A_1)=\left(-A_2\frac{\partial^2}{\partial A_2^2}-
\frac{\partial}{\partial A_2}\right)\d(A_2-A_1).
\eeq
Dividing $\hat H''_{kin}$ by the entropy factor $1/\sqrt{A_2}$, we find the kinetic part of the auxiliary
Hamiltonian (\ref{H-aux}), namely,
\beq
\hat H'_{kin}=-A_2^{\frac{3}{2}}\frac{\partial^2}{\partial A_2^2}
-A_2^{\frac{1}{2}}\frac{\partial}{\partial A_2} .
\eeq
Finally, by use of (\ref{H-inversion}) (where $\NN(A)$ has to be replaced by the subleading term
of (\ref{entropy-A}) only, because the exponential part is absorbed in the critical value of the
boundary cosmological constants), we find the kinetic term of the final Hamiltonian (\ref{H}).

As we have demonstrated in this appendix, the convergence problems encountered
in the two methods are complementary, in the sense that
the only problem with the auxiliary Hamiltonian method arises in the cosmological term,
which instead presents no problems with regard to the fractional derivative.

\bibliographystyle{unsrt}

\end{document}